\documentclass[pra,twocolumn,showpacs]{revtex4}
\usepackage{fancyhdr}
\usepackage{amssymb}
\usepackage{amsmath}
\usepackage[mathscr]{eucal}
\usepackage{latexsym}
\usepackage{amsbsy}
\usepackage{float}
\usepackage[dvips]{graphicx}
\usepackage{epsfig}
\usepackage{subfigure}
\usepackage{wrapfig}

\newcommand{\be}{\begin{equation}}
\newcommand{\ee}{\end{equation}}
\newcommand{\bes}{\begin{equation*}}
\newcommand{\ees}{\end{equation*}}
\newcommand{\bea}{\begin{eqnarray}}
\newcommand{\eea}{\end{eqnarray}}
\newcommand{\bean}{\begin{eqnarray*}}
\newcommand{\eean}{\end{eqnarray*}}
\newcommand{\ba}{\begin{array}}
\newcommand{\ea}{\end{array}}

\newcommand{\vk}{{\bf k}}
\newcommand{\vq}{{\bf q}}
\begin{document}


\title{Tuning Rashba and Dresselhaus spin-orbit couplings: \\effects 
on singlet and triplet condensation with Fermi atoms}

\author{L. Dell'Anna, G. Mazzarella, and L. Salasnich}

\affiliation{Dipartimento di Fisica ``Galileo Galilei'' and CNISM,
Universit\`a di Padova, Italy}

\begin{abstract}
We investigate the pair condensation of a two-spin-component Fermi gas
in the presence of both Rashba and Dresselhaus spin-orbit couplings.
We calculate the condensate fraction in the BCS-BEC crossover
both in two and in three dimensions by taking into account
singlet and triplet pairings. These quantities are studied
by varying the spin-orbit interaction from the case with the only Rashba to
the equal-Rashba-Dresselhaus one. We find that, by mixing
the two couplings, the singlet pairing decreases while
the triplet pairing is suppressed in the BCS regime and increased in the BEC
regime, both in two and three dimensions.
At fixed spin-orbital strength, the greatest total condensate fraction is
obtained when only one coupling (only Rashba or only Dresselhaus) is present.
\end{abstract}


\maketitle

\section{Introduction}

In the last years the predicted crossover \cite{eagles,leggett,noziers}
from the Bardeen-Cooper-Schrieffer (BCS) state of weakly bound Fermi
pairs to the Bose-Einstein condensate (BEC) of molecular dimers has
been observed by several experimental groups
\cite{greiner,regal,kinast,zwierlein,chin,ueda}. In particular,
three seminal experiments with two hyperfine
component Fermi vapours of $^{40}$K atoms \cite{regal}
or $^6$Li atoms \cite{zwierlein,ueda}
in the BCS-BEC crossover, have
been performed to study the condensate fraction
of Cooper pairs \cite{yang}, which is directly related
to the off-diagonal-long-range order of the two-body density
matrix of fermions \cite{penrose,campbell}. At very low temperature
the experimental results with $^6$Li atoms \cite{zwierlein,ueda}
show an excellent agreement with
the zero-temperature theoretical predictions
of mean-field approaches \cite{sala-odlro,ortiz} and Monte-Carlo simulations
\cite{astrakharchik}. However, as discussed in \cite{ohashi},
when the effects of temperature cannot be neglected
it is necessary to include beyond mean-field corrections
to reproduce quantitatively the experimental data.
Intensive theoretical studies have been developed
on the condensate fraction along BCS-BEC crossover for a
two-dimensional (2D) Fermi gas \cite{marini,sala-odlro2,conduit,he0,du,parish},
and for a three-spin-component Fermi gas with SU(3)
symmetry \cite{desilva,sala-odlro3,du2} as well.
The recent experimental realization of 2D degenerate Fermi gases
for ultra-cold atoms in a highly anisotropic disk-shaped potential
\cite{turlapov} is one of the reasons of the growing interest
for fermions in reduced dimensionality.

Artificial spin-orbit coupling has been recently realized
in neutral bosonic systems \cite{spielman}.
In such systems the strength of the coupling can be optically tuned
and this is indeed a useful tool also for ultracold
fermions \cite{dalibard,chapman}. These achievements have stimulated
theoretical efforts in understanding the spin-orbit effects
with Rashba \cite{rashba} and Dresselhaus \cite{dress} terms
in the BCS-BEC crossover \cite{shenoy,shenoy2,gong,
hu,yu,iskin,yi,dms,iskin2,zhou,jiang,sa,chen,zhou2,yangwan,iskin3,sa2,he}.
The evolution from BCS to BEC superfluidity was intensively studied
in the presence of spin-orbit coupling for a 3D uniform Fermi
gas \cite{shenoy2,gong,hu,yu,iskin,yi,dms,iskin2,jiang,sa,zhou2} and in the
2D case \cite{dms,zhou,chen,zhou2,yangwan,he}.
In Ref. \cite{dms} we have analyzed the effects of the spin-orbit coupling
on the condensate fraction by studying both the singlet and the
triplet pairing contributions in the presence of the Rashba coupling.
{We stress that very recently a theoretical proposal by Liu and co-workers \cite{liu} has been implemented
in two experiments \cite{ne1,ne2} to observe the spin-orbit coupling effects 
on atomic Fermi gases with the Rashba term equal to the Dresselhaus one.

Motivated by such realizations in the lab}, in this paper we extend and complete the zero-temperature
study presented in \cite{dms}, where it was detailed analyzed
only the case without Dresselhaus coupling.
In particular we investigate the condensate fraction along
the BCS-BEC crossover with Rashba and Dresselhaus spin-orbit couplings
both in 2D and in 3D analyzing singlet and triplet contributions
to pairing condensation. We study these quantities at zero temperature
by gradually including the effect
of the Dresselhaus term in the spin-orbit coupling.
We show that along this tuning the singlet contribution
to the condensate fraction always decreases, while the triplet contribution
is strongly suppressed in the BCS regime and enhanced in the BEC one,
both in 2D and in 3D. Indeed this enhancement in 2D
takes place when the binding energy is greater
than the Fermi energy and in 3D when the dimensionless
interaction parameter $y=1/(k_F a_s)$ ($k_F$ is the Fermi linear
momentum and $a_s$ the interatomic s-wave scattering length) is mainly
positive. We find that the total condensate fraction is greater when only the
Rashba coupling is active. Instead, when the Rashba and
Dresselhaus couplings are equal
the total condensate fraction is the same as that obtained
in the absence of spin-orbit coupling.
The chemical potential and the pairing gap decrease in both the two regimes,
when the coupling is changed from the only-Rashba to
equal-Rashba-Dresselhaus case. Our theoretical predictions
on the effects of Rashba and Dresselhaus couplings
can be experimentally tested. In particular, we suggest that
the condensate fraction of singlet and triplet pairs
is detectable by {suitably} extending the procedure used
in previous experiments \cite{zwierlein,ueda}.

\section{The model}

We describe a gas of two-spin-component Fermi atoms with spin-orbit
couplings \cite{dress,rashba} by using the following one-body Hamiltonian
\bea
\label{H0}
H_0&=&\sum_{\vk}\psi(\vk)^\dagger\Big\{\frac{\hbar^2 k^2}{2m}-\mu
+\hbar\big[v_R(\sigma_x k_y-\sigma_y k_x)
\nonumber
\\
&+&v_D(\sigma_x k_y+\sigma_y k_x) \big]\Big\}\psi(\vk)\;,
\eea
where $\mu$ is the chemical potential and $v_D$ and $v_R$ are,
respectively, the Rashba and Dresselhaus
velocities; $\sigma_x$ and $\sigma_y$ denote the Pauli matrices
in the $x$ and $y$ directions, and
$\psi(\vk)=(\psi_{\uparrow}(\vk),\psi_{\downarrow}(\vk))^T$
is the Nambu spinor.
Notice that we can rewrite Eq. (\ref{H0}) as
\be
\label{H0_A}
H_0=\sum_{\vk}\psi(\vk)^\dagger\left[\frac{\hbar^2}{2m}
\left(\vec k+\vec A\right)^2-(\mu+mv^2)\right]
\psi(\vk)
\;,
\ee
where
\bea
v_R&=&v\cos\theta
\\
v_D&=&v\sin\theta
\\
\vec A&=&\frac{mv}{\hbar}\left(\ba{c}
(\sin\theta-\cos\theta)\sigma_y\\
(\sin\theta+\cos\theta)\sigma_x\\
0
\ea\right)\; .
\eea
In addition to the one-body Hamiltonian $H_0$ we consider
the two-body interaction Hamiltonian given by
\be
H_I=-\frac{g}{V} \sum_{\vk \vk' \vq}\psi_{\uparrow}^\dagger(\vk+\vq)
\psi_{\downarrow}^\dagger(-\vk)
\psi_{\downarrow}(-\vk'+\vq)\psi_{\uparrow}(\vk') \;,
\label{zorro}
\ee
where $g>0$, which corresponds to attractive interaction.
The total Hamiltonian thus reads
\be
H = H_0 + H_I \; .
\label{totH}
\ee
From this Hamiltonian we calculate the effects
of Rashba and Dresselhaus spin-orbit couplings
on singlet and triplet condensation with Fermi atoms
in the full BCS-BEC crossover.

\section{Gap order parameter and condensates}

By decoupling at the mean-field level the two-body
interaction (\ref{zorro}) we get
\be
\label{mf}
H_I=V\frac{|\Delta|^2}{g}-
\sum_\vk\left(\Delta^*\psi_\downarrow(-\vk)\psi_\uparrow(\vk)+\Delta
\psi_\uparrow^\dagger(\vk) \psi_\downarrow^\dagger(-\vk)\right)\;,
\ee
where $V$ is the volume and
\be
\Delta=(g/V)\sum_\vk \langle \psi_\downarrow(-\vk)
\psi_\uparrow(\vk)\rangle
\ee
is the familiar gap order parameter describing the correlation
energy of singlet Cooper pairs.

As discussed in \cite{dms},
from the Hamiltonian (\ref{totH}) with $H_0$ given by Eq. (\ref{H0_A}) and
$H_I$ by Eq. (\ref{mf}) we can calculate the spectrum of single-particle
elementary excitations, which is given by
\bea
E_1(\vk)&=&\sqrt{(\xi_\vk-|\gamma(\vk)|)^2
+|\Delta|^2}
\\
E_2(\vk)
&=&\sqrt{(\xi_\vk+|\gamma(\vk)|)^2+|\Delta|^2}
\\
E_3(\vk)&=&-E_1(\vk)
\\
E_4(\vk)&=&-E_2(\vk)
\eea
with
\be
\gamma(\vk)=\hbar v_R(k_y+ik_x)+\hbar v_D(k_y-ik_x)
\ee
and
\be
\xi_{\vk}=\hbar^2k^2/2m-\mu \; .
\ee
Moreover, the number of particles reads
\be
\label{number}
N=\sum_{\vk}\Big\{1-\frac{\xi_{\vk}-
|\gamma(\vk)|}{2E_1(\vk)}
-\frac{\xi_{\vk}+
|\gamma(\vk)|}{2E_2(\vk)}\Big\} \;,
\ee
while the energy gap $\Delta$ is obtained by solving
the corresponding gap equation
\be
\label{gap}
\frac{V}{g}=\frac{1}{4}\sum_\vk\left(\frac{1}{E_1(\vk)}
+\frac{1}{E_2(\vk)}\right) \; .
\ee
In addition, the condensate number $N_c$ \cite{leggett2} of Cooper pairs
is given by
\be
N_c=N_{0}+N_{1},
\ee
where
\bea
\label{n0}
&&N_{0}=\sum_\vk \left|\langle \psi_\uparrow(\vk)\psi_\downarrow(-\vk)
\rangle\right|^2
\\
\nonumber&&= \frac{|\Delta|^2}{16}
\sum_\vk\left(\frac{1}
{E_1(\vk)}+\frac{1}{E_2(\vk)}\right)^2
\eea
is the singlet, with total spin $0$, contribution to the condensate, whereas
\bea
\label{n1}
&&N_{1}=\sum_\vk \left|\langle \psi_\uparrow(\vk)\psi_\uparrow(-\vk)
\rangle\right|^2
\\
\nonumber&&= \frac{|\Delta|^2}{16}
\sum_\vk\left(\frac{1}
{E_1(\vk)}-\frac{1}{E_2(\vk)}\right)^2
\eea
is the triplet one, with total spin $1$.
Eqs. (\ref{number}), (\ref{gap}), (\ref{n0}) and (\ref{n1}) are
the starting point of our present investigation.
The finite-temperature version of these equations
can be found in our previous paper \cite{dms}.
It is important to notice that, even if at the mean-field level only the singlet energy gap $\Delta=\Delta_{\downarrow\uparrow}$ appears, while the
triplet one, $\Delta_{\uparrow\uparrow}=
(g/V)\sum_\vk \langle \psi_\uparrow(-\vk)
\psi_\uparrow(\vk)\rangle$, is absent, as one can see from Eq. (\ref{zorro}),
triplet pairing can be generated by the presence of the spin-orbital
interaction.

We study our system both in two dimensional (2D) and in three dimensional
(3D) case in the absence of the temperature, $T=0$. With this purpose,
we have to analyze the key quantities for the 2D case - the binding energy
$\epsilon_B$, the condensate densities $n_S=N_S/V$ ($S=0$ for the singlet and $S=1$
for the triplet), the chemical potential $\mu$, and the gap $\Delta$ -
and for the 3D case - the interaction parameter $y$, the condensate
densities $n_S$, the chemical potential $\mu$, and the gap $\Delta$.
We follow the same path as in \cite{dms}, that is we define the
dimensionless parameters
\bea
&& x_0=\frac{\mu}{\Delta}\\
&& x_1=2m \frac{(v_R-v_D)^2}{\Delta}=\frac{2mv^2}{\Delta}(1-\sin 2\theta)\\
&& x_2=2m \frac{(v_R+v_D)^2}{\Delta} =\frac{2mv^2}{\Delta}(1+\sin 2\theta)\;
\eea
so that for the 2D case we have
\bea
&&\frac{\epsilon_B}{\Delta}=\lim_{\Lambda\rightarrow \infty}
\frac{2\Lambda^2}{\exp[I_g(x_0,x_1,x_2)/\pi]-1} \\
&&\frac{2n_{S}}{n}=\frac{1}{8}\frac{I^{2}_{N_{S}}(x_0,x_1,x_2)}
{I^{2}_N(x_0,x_1,x_2)}\\
&&\frac{\mu}{\epsilon_F}=\frac{2\pi x_0}{I^{2}_N(x_0,x_1,x_2)}\\
&&\frac{\Delta}{\epsilon_F}=\frac{2\pi}{I^{2}_N(x_0,x_1,x_2)}\\
&&\frac{(v_R\mp v_D)^2}{v_F^2}=\frac{\pi}{2}\frac{x_{1,2}}{
I^{2}_N(x_0,x_1,x_2)}
\eea
with $\Lambda$ the ultraviolet momentum cut-off and $n=N/V$ the particle 
density. For the 3D case, one has
\bea
&&y\equiv \frac{1}{k_F a_s}=\frac{1}{3^{1/3}\pi^{5/3}}\frac{I_{a_s}
(x_0,x_1,x_2)}{I^{3}_N(x_0,x_1,x_2)^{1/3}}\\
&&\frac{2n_{S}}{n}=\frac{1}{8}\frac{I^{3}_{N_{S}}(x_0,x_1,x_2)}
{I^{3}_N(x_0,x_1,x_2)}\\
&&\frac{\mu}{\epsilon_F}=4\left(\frac{\pi}{3}\right)^{2/3} x_0
I^{3}_{N}(x_0,x_1,x_2)^{-2/3}\\
&&\frac{\Delta}{\epsilon_F}=4\left(\frac{\pi}{3}\right)^{2/3}
I^{3}_{N}(x_0,x_1,x_2)^{-2/3}\\
&&\frac{(v_R\mp v_D)^2}{v_F^2}=\left(\frac{\pi}{3}\right)^{2/3}x_{1,2}\,
I^{3}_N(x_0,x_1,x_2)^{-2/3}
\eea
with $k_F$ the Fermi wave vector and $a_s$
the interatomic s-wave scattering length.
All the above equations are written in terms of
the following dimensionless integrals:
\begin{widetext}
\bea
&&I_g(x_0,x_1,x_2)=\frac{1}{2}\int^\Lambda d^2{\vq}\sum_{r=\pm 1}
\frac{1}{\sqrt{\left(q^2-x_0+r\sqrt{x_1 q_x^2+x_2 q_y^2}\right)^2+1}}\nonumber\\
&&I^{d}_{N}(x_0,x_1,x_2)=\int d^d{\vq}
\left(1-\frac{1}{2}\sum_{r=\pm 1}
\frac{q^2-x_0+r\sqrt{x_1 q_x^2+x_2 q_y^2}}{\sqrt{\left(q^2-x_0+
r\sqrt{x_1 q_x^2+x_2 q_y^2}\right)^2+1}}\right)\nonumber\\
&&I^{d}_{N_{S}}(x_0,x_1,x_2)=\int d^d{\vq}
\left(\sum_{r=\pm 1}\frac{r^S}
{\sqrt{\left(q^2-x_0+r\sqrt{x_1 q_x^2+x_2 q_y^2}\right)^2+1}}\right)^2\nonumber\\
&&I_{a_s}(x_0,x_1,x_2)=\int d^3{\vq} \left(\frac{1}{q^2}-\frac{1}{2}\sum_{r=\pm 1}
\frac{1}{\sqrt{\left(q^2-x_0+r\sqrt{x_1 q_x^2+x_2 q_y^2}\right)^2+1}}
\right) \;,
\eea
\end{widetext}
where $d=2$ for 2D case, while $d=3$ for 3D; $S=0$ for the
singlet and $S=1$ for the triplet contribution.

\section{Results}

\begin{figure}[ht]
\includegraphics[width=6.6cm]{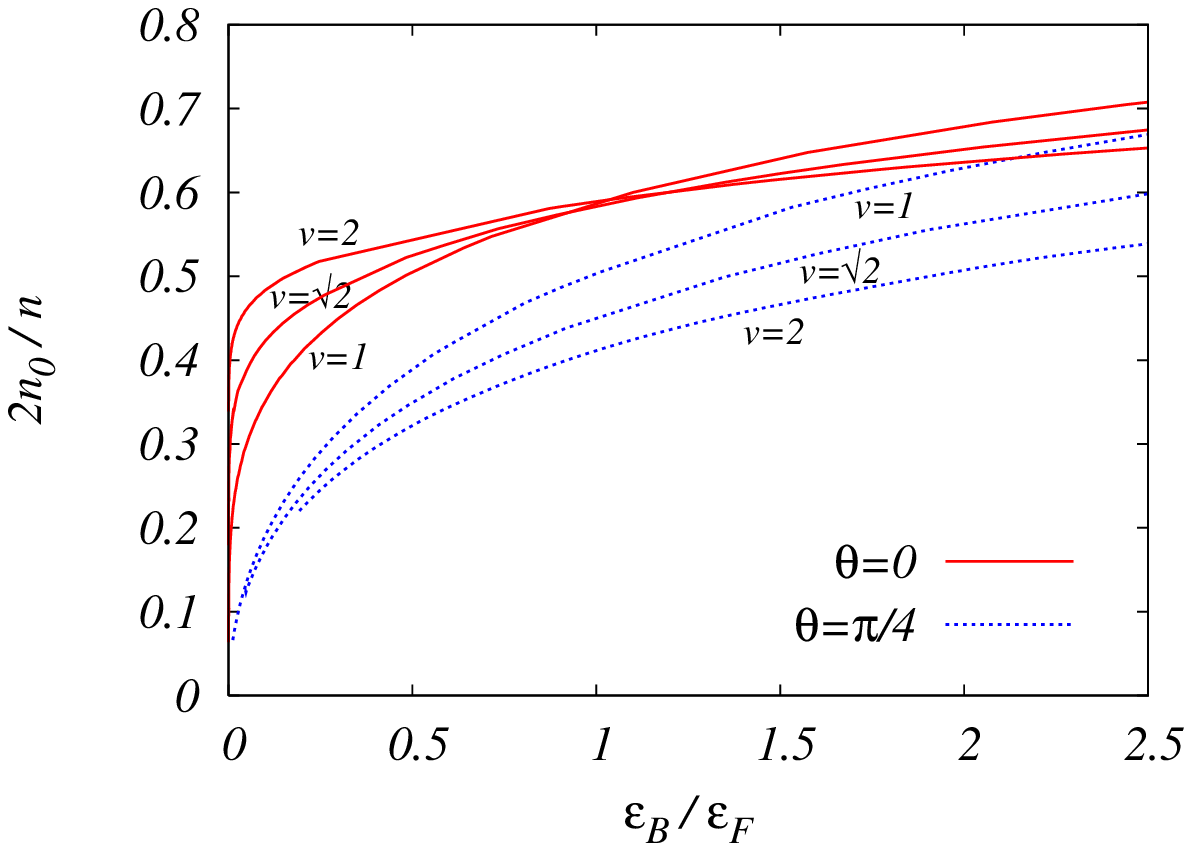}
\includegraphics[width=6.6cm]{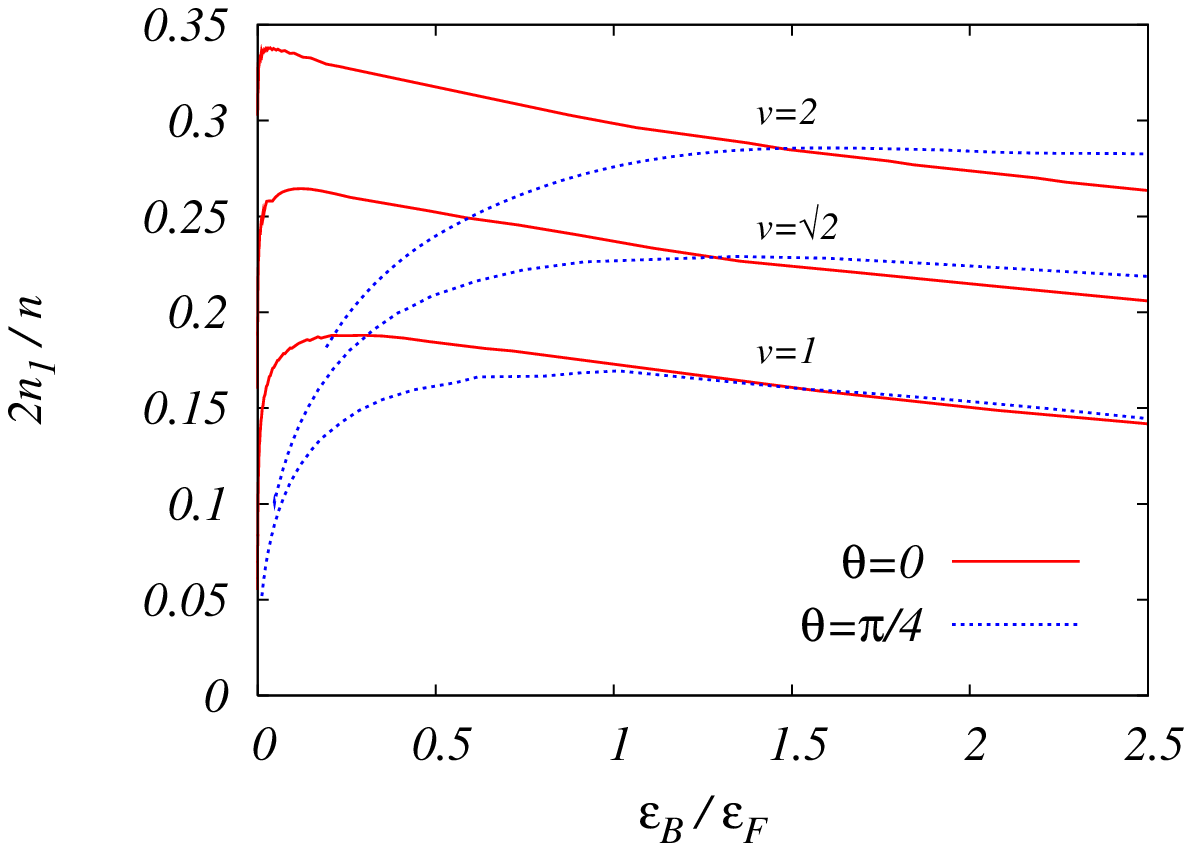}
\includegraphics[width=6.6cm]{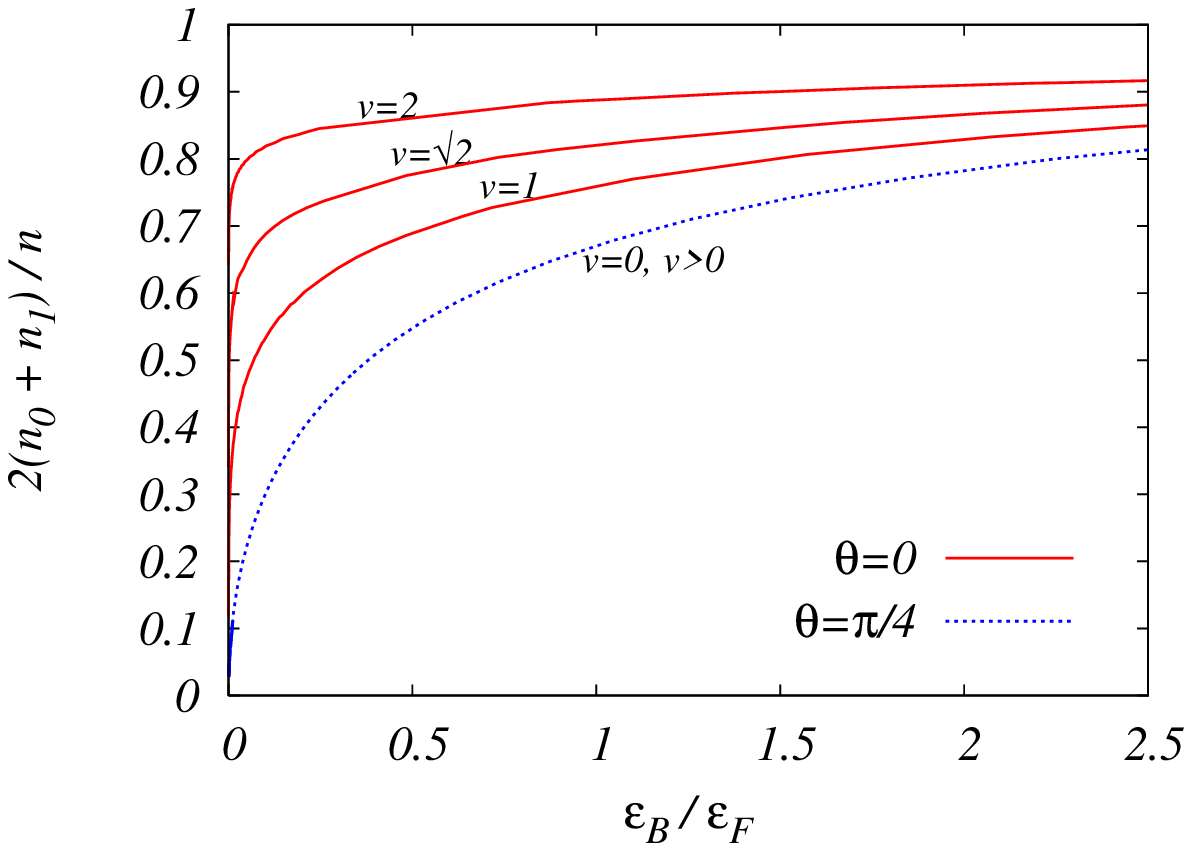}
\caption{(Color online) 2D Fermi superfluid.
Singlet (top panel) and triplet (middle panel)
contributions to the condensate fraction for $v^2=1,2,4$
(in units of $v_F^2$) and
for $\theta=0$, i.e. only Rashba term, (red solid curves) and
$\theta=\pi/4$, i.e. for equal Rashba and Dresselhaus terms,
(blue dashed line). Bottom panel: the total condensate fraction,
i.e. the sum of both singlet and triplet contributions, for the
same values of $v$ and $\theta$. Notice that for $\theta=\pi/4$
(blue dashed line), the full condensate fraction is the same as
without spin-orbit ($v=0$).}
\label{fig.1}
\end{figure}

In Fig. \ref{fig.1} we plot the condensate fraction
of the 2D Fermi system as a function of the scaled
binding energy $\epsilon_B/\epsilon_F$, with $\epsilon_F$ the 2D
Fermi energy, for different values of the characteristic velocity
$v = \sqrt{v_R^2+v_D^2}$ and two values of the mixing angle
$\theta = \arctan{\left({v_D/v_R}\right)}$.
In the first two panels there are the
singlet (top panel) and triplet (middle panel)
contributions to the condensate fraction, while
in the lower panel there is the total condensate fraction.
The results are shown for $\theta=0$ ($v_D=0$, solid curves)
and $\theta=\pi/4$ ($v_R=v_D$, dashed curves)
and three values of the velocity $v$.

\begin{figure}[ht]
\includegraphics[width=6.6cm]{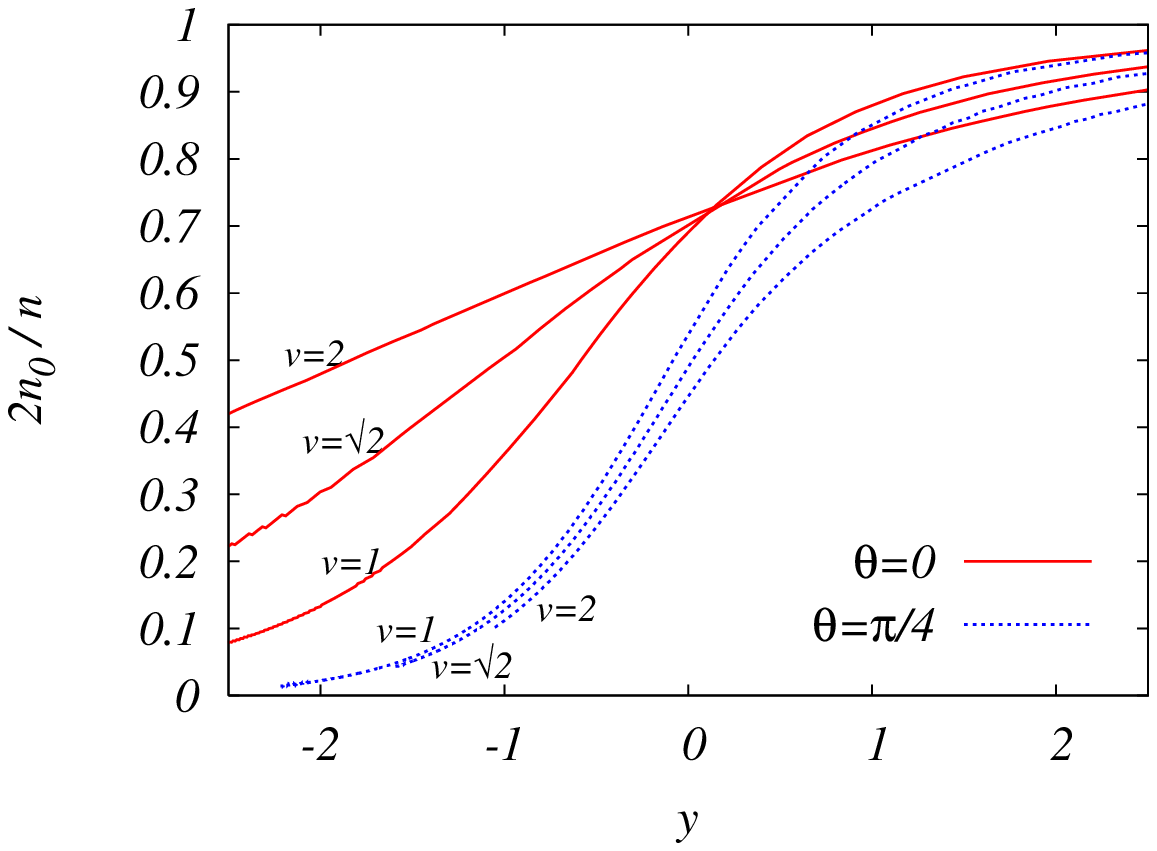}
\includegraphics[width=6.6cm]{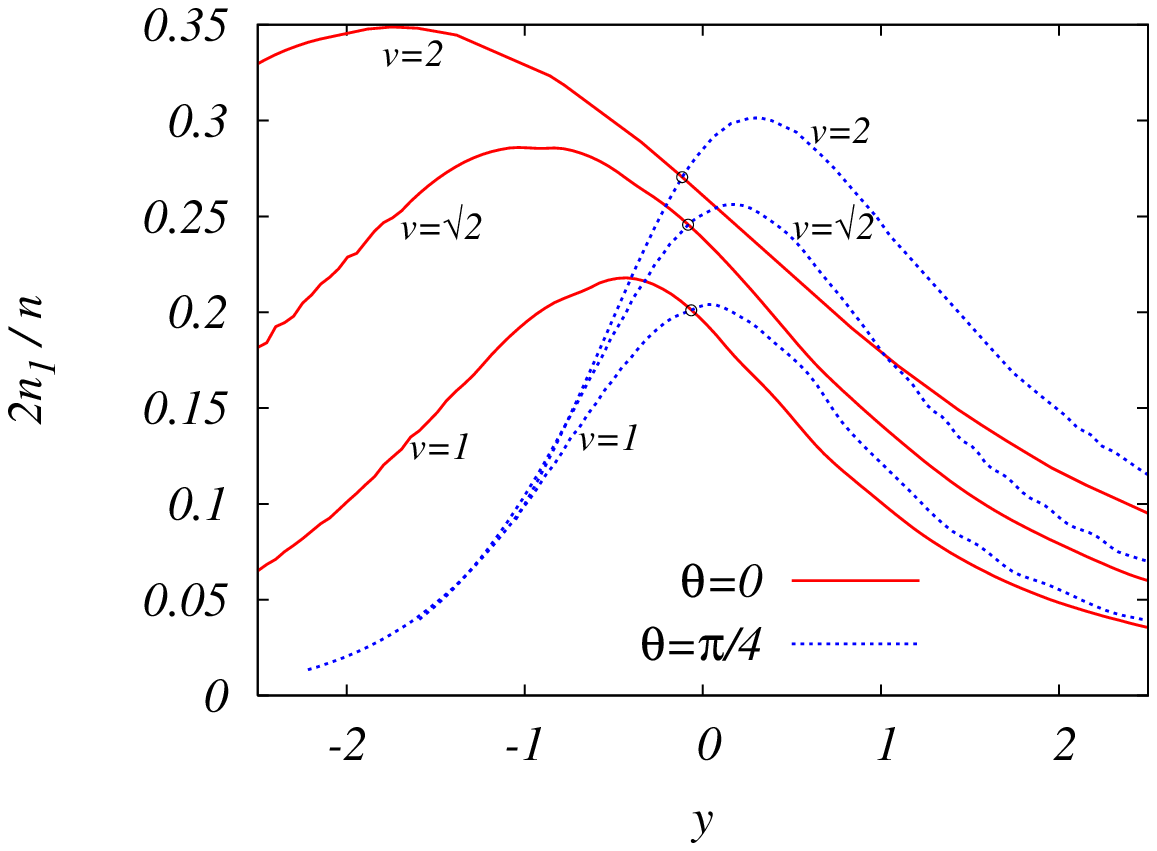}
\includegraphics[width=6.6cm]{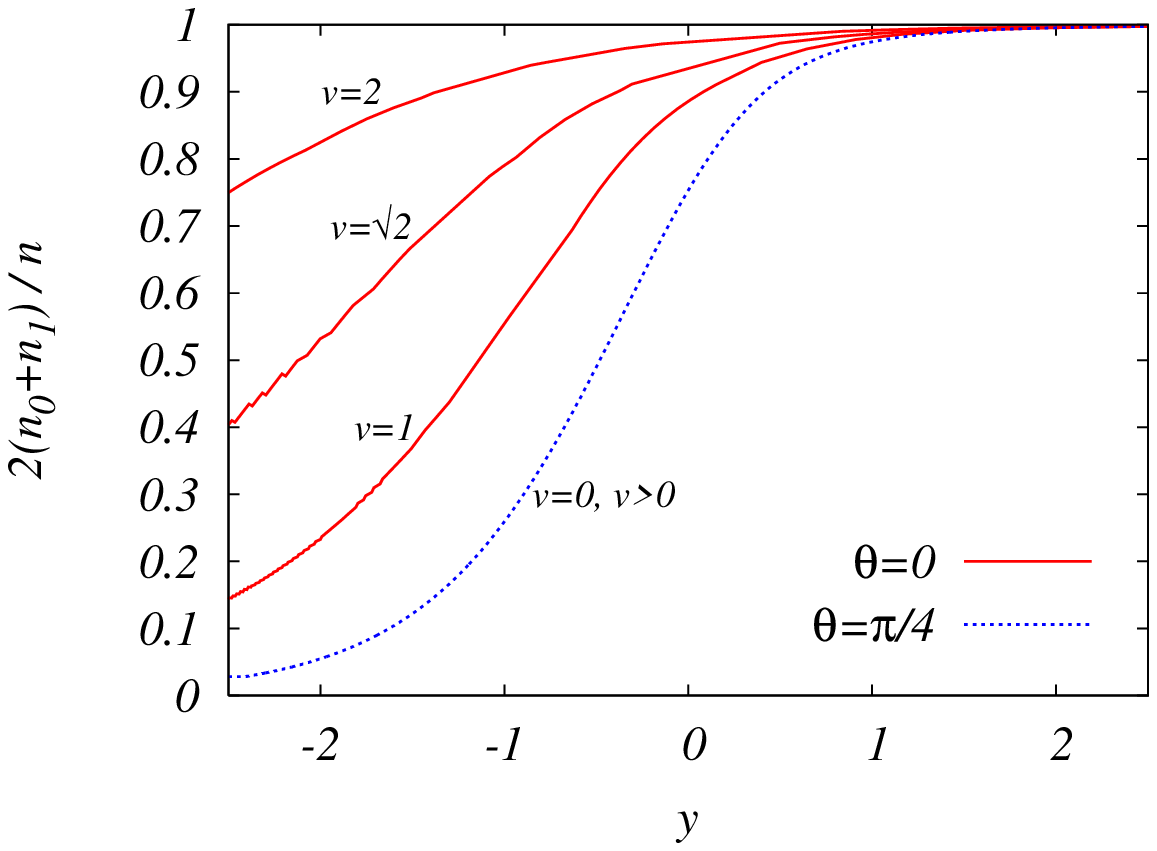}
\caption{(Color online) 3D Fermi superfluid. Singlet (top panel) and triplet
(middle panel) contributions to the
condensate fraction for $v^2=1,2,4$
(in units of $v_F^2$) and for $\theta=0$, i.e. only Rashba
term, (red solid curves) and $\theta=\pi/4$, i.e. for equal Rashba and
Dresselhaus terms, (blue dashed line). Bottom panel: the total condensate
fraction, i.e. the sum of both singlet and triplet contributions,
for the same values of $v$ and $\theta$. Notice that for
$\theta=\pi/4$ (blue dashed line), the full condensate fraction
is the same as without spin-orbit ($v=0$).}
\label{fig.2}
\end{figure}

In Fig. \ref{fig.2} we plot instead the condensate fraction
of the 3D Fermi system as a function of the scaled
interaction strength $y=1/(k_Fa_s)$ with $k_F$ the 3D
Fermi linear momentum and $a_s$ the s-wave scattering length.
As in Fig. \ref{fig.1} in the first two panels there are the
singlet (top panel) and triplet (middle panel)
contributions to the condensate fraction, while
in the lower panel there is the total condensate fraction.
Again, results are shown for $\theta=0$ ($v_D=0$, solid curves)
and $\theta=\pi/4$ ($v_R=v_D$, dashed curves)
and three values of the velocity $v$. The two sets of plots corresponding to
2D and 3D share many features and would have looked very similar if we had
plotted Fig. \ref{fig.1} in terms of
$\ln(1/k_F a_{2D})\equiv\ln(\epsilon_B/2\epsilon_F)/2$.

\begin{figure}[ht]
\includegraphics[width=7cm]{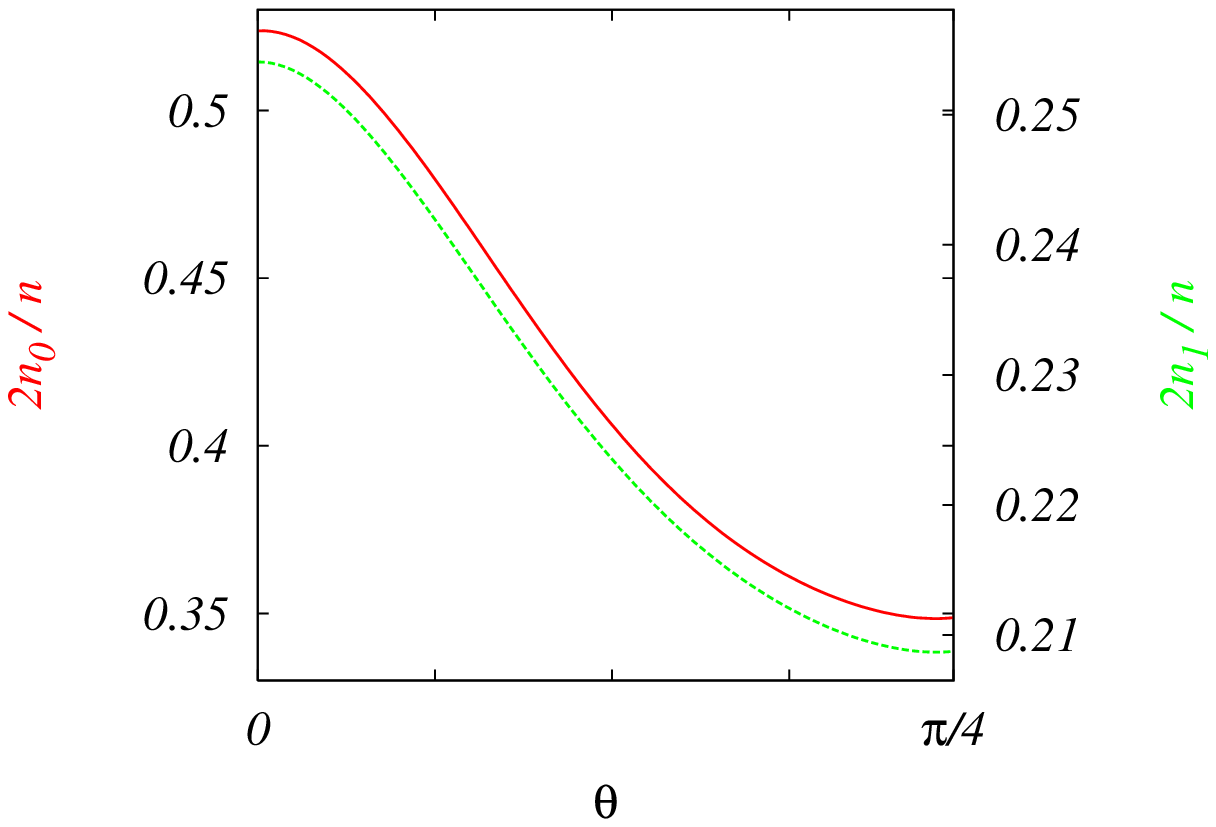}
\includegraphics[width=4.cm,height=4cm]{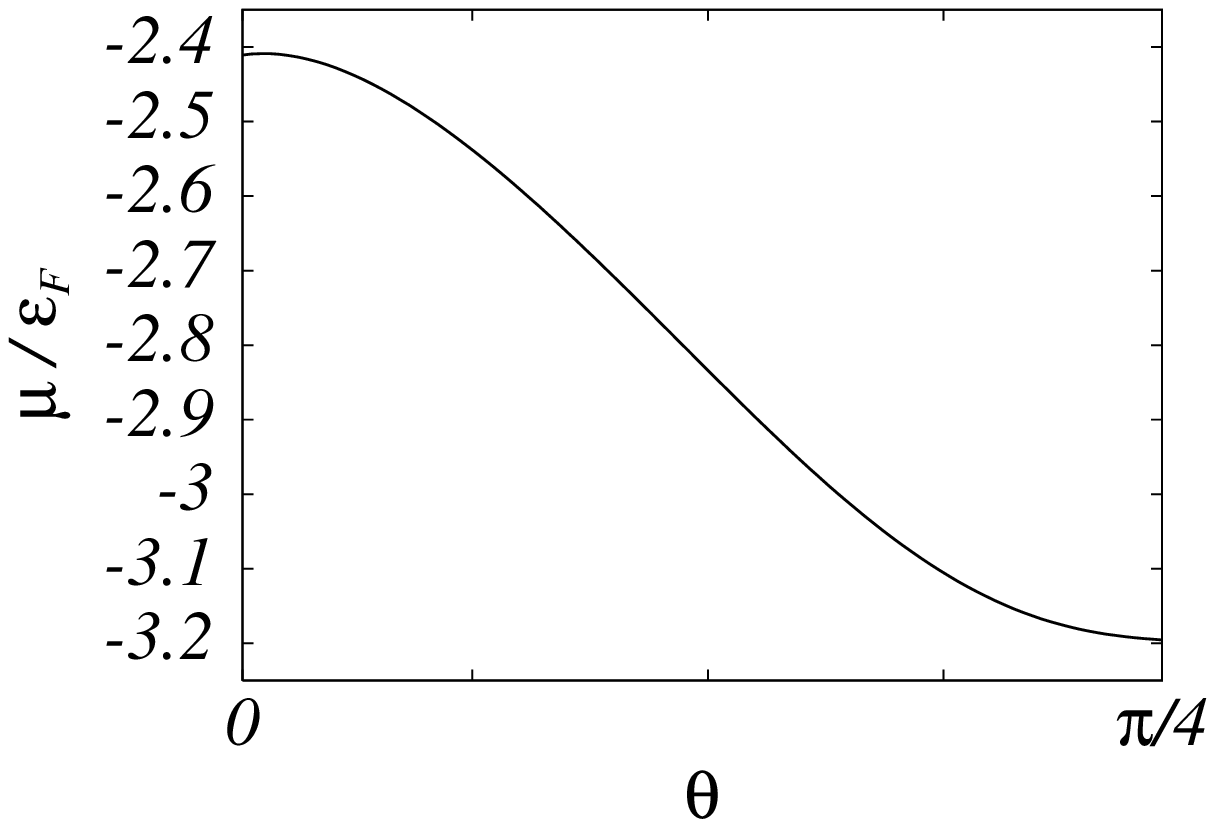}
\includegraphics[width=4.cm,height=4cm]{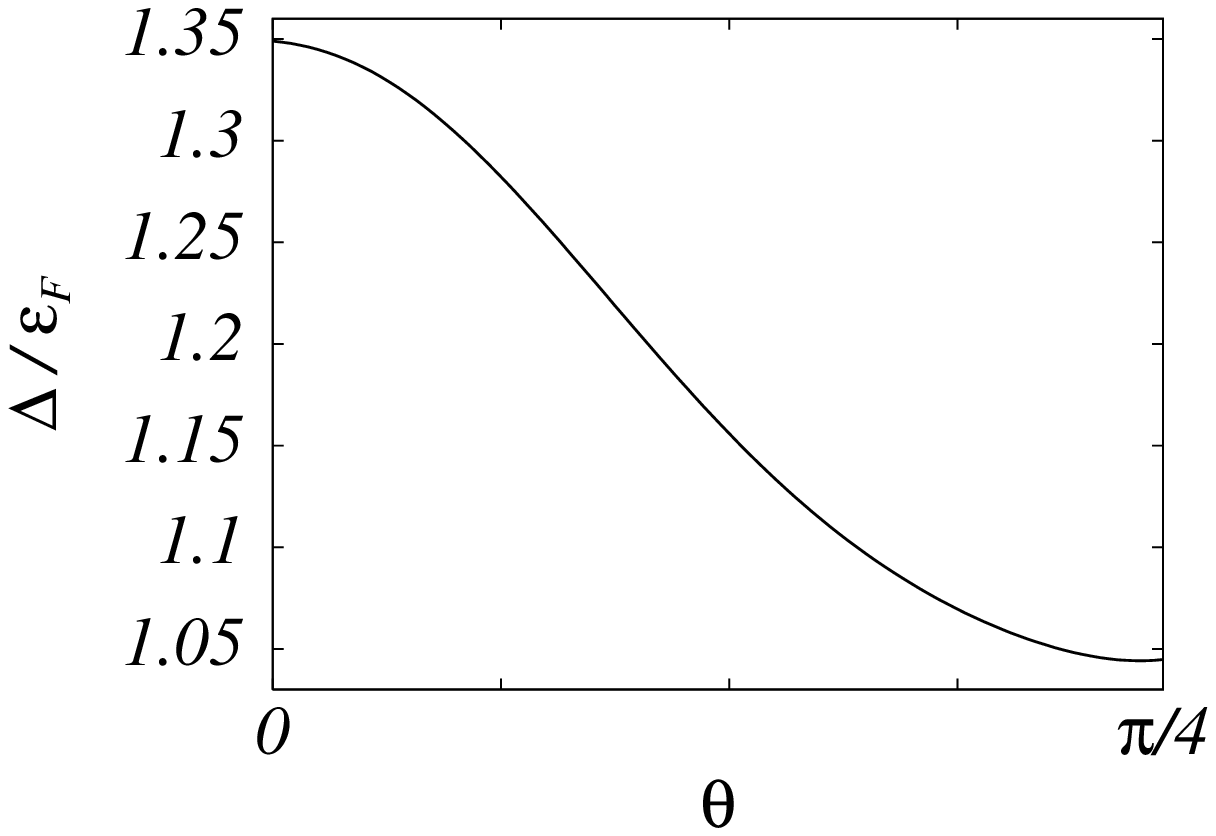}
\caption{(Color online) 2D Fermi superfluid.
Top panel: Singlet (red solid line, left y-axis)
and triplet (green dashed line, right y-axis) contributions to the
condensate fraction, for $v^2=2v_F^2$ and
$\epsilon_B=0.5 \epsilon_F$, as functions of $\theta$. Notice that
$2n_0/n$ and $2n_1/n$ are plotted with different scales.
Left bottom panel: chemical potential, $\mu$ (in units of $\epsilon_F$),
for $v^2=2v_F^2$ and $\epsilon_B=0.5
\epsilon_F$, as a function of $\theta$.
Right bottom panel: gap function, $\Delta$ (in units of $\epsilon_F$),
for $v^2=2v_F^2$ and $\epsilon_B=0.5
\epsilon_F$, as a function of $\theta$. }
\label{fig.3}
\end{figure}

Our calculations show that the singlet contribution to the condensate
fraction, $2n_0/n$, both in 2D and in 3D, decreases when one fixes $v/v_F$
and moves from the only-Rashba (or only-Dresselhaus) case ($\theta=0$)
to equal-Rashba-Dresselhaus case ($\theta=\pi/4$), as it can be observed
from the top panels of Figs. \ref{fig.1} and \ref{fig.2}. In the case of only
Rashba ($\theta=0$) we observed \cite{dms} that,
for one value of the scattering parameter close to the crossover
($\epsilon_B\simeq \epsilon_F$ in 2D and $y\simeq 0$ in 3D)
and for $v\gtrsim v_F$, $n_0/n$ does not depends on $v$, namely there is a
nodal point for the singlet condensate fraction when one increases
largely enough the Rashba coupling, see red solid curves in the
top panels of Figs. \ref{fig.1} and \ref{fig.2}.
Regarding the triplet contribution, $2n_1/n$, we find that
for sufficiently small values of the binding energy
$\epsilon_B/\epsilon_F$ (in 2D) or for $y<0$ (in 3D),
$2n_1/n$ is suppressed by mixing the two spin-orbital couplings,
whereas when $\epsilon_B/\epsilon_F$ (in 2D) becomes
large enough ($\epsilon_B\gtrsim 1.5\epsilon_F$) and
$y\gtrsim 0$ (in 3D), the triplet contribution is enhanced, see the
middle panels of Figs. \ref{fig.1} and \ref{fig.2}.
The full condensate fraction, $n_c/n=2(n_0+n_1)/n$, at fixed $v/v_F$, is
maximum for only-Rashba (or only-Dresselhaus) case ($\theta=0$).
In the extreme case of equal Rashba and Dresselhaus contributions
($\theta=\pi/4$), instead, the total condensate fraction is the
same as that obtained without spin-orbit at all, i.e. with $v=0$,
see the dotted line
of the bottom panels of Figs. \ref{fig.1} and \ref{fig.2}.
In particular, for the case of $\theta=\pi/4$  the following results
for the condensate fraction, the chemical potential and the gap function hold
\bea
\label{erd_nc}
&&\frac{n_c}{n}(v,\theta=\pi/4)=\frac{n_c}{n}(v=0)\;,\\
\label{erd_mu}
&&\frac{\mu}{\epsilon_F}(v,\theta=\pi/4)=\frac{\mu}{\epsilon_F}(v=0)-\frac{2v^2}{v_F^2}\;,\\
\label{erd_del}
&&\Delta(v,\theta=\pi/4)=\Delta(v=0)\;.
\eea
The chemical potential is consistent with the known perturbative
result for small $v$ and small $\epsilon_B$, which is $\mu\simeq
\epsilon_F-mv^2$. In the case of $\theta=\pi/4$, Eq.(\ref{erd_mu})
is valid for all values of the binding energy $\epsilon_B$ and comes simply
from gauge transforming the fields,
$\psi({\bf r})\rightarrow e^{iyA_y}\psi({\bf r})$,
in the Hamiltonian Eq. (\ref{H0_A}).

Let us now investigate in detail the condensate fractions $n_0/n$
and $n_1/n$, the chemical potential $\mu$ and the energy gap $\Delta$
when $\theta$ is increased from $0$ to $\pi/4$.
For computational simplicity, we consider the two dimensional case.

\begin{figure}[ht]
\includegraphics[width=7cm]{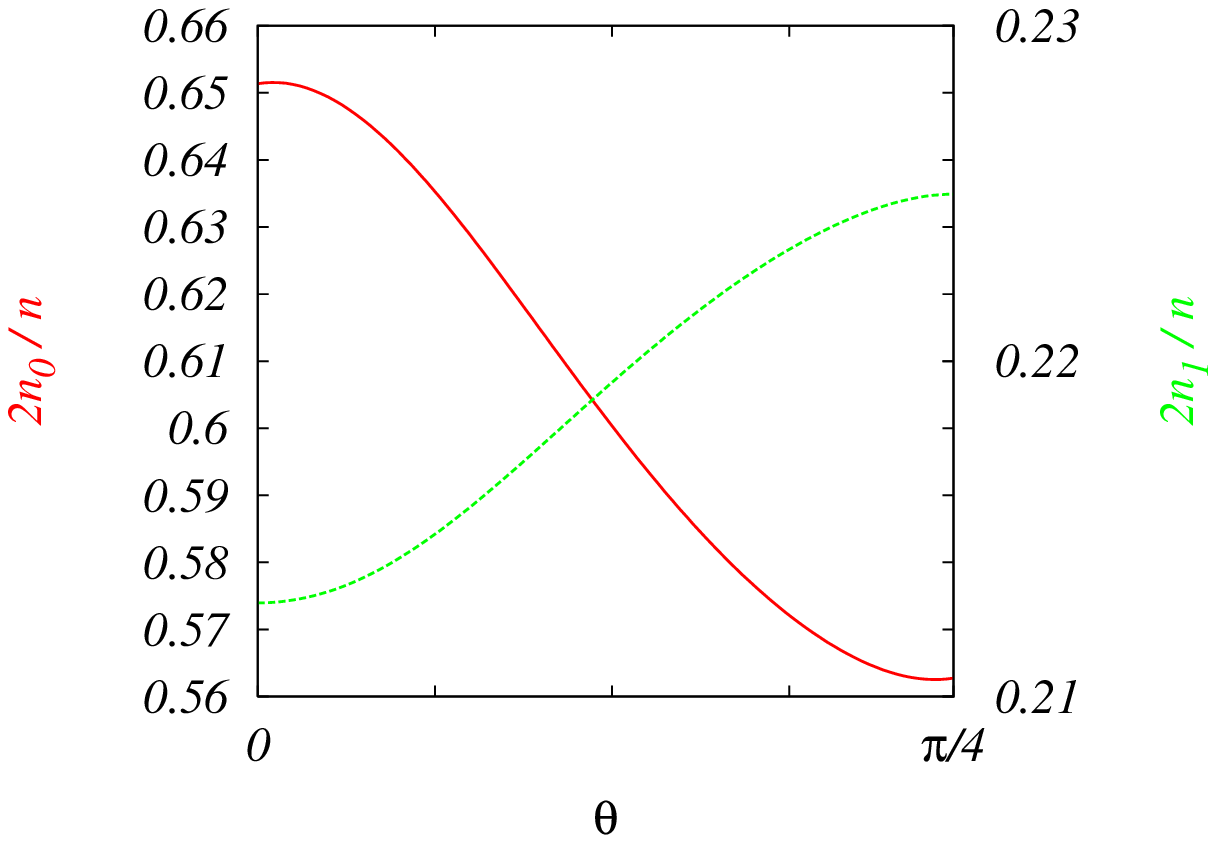}
\includegraphics[width=4.cm,height=4.cm]{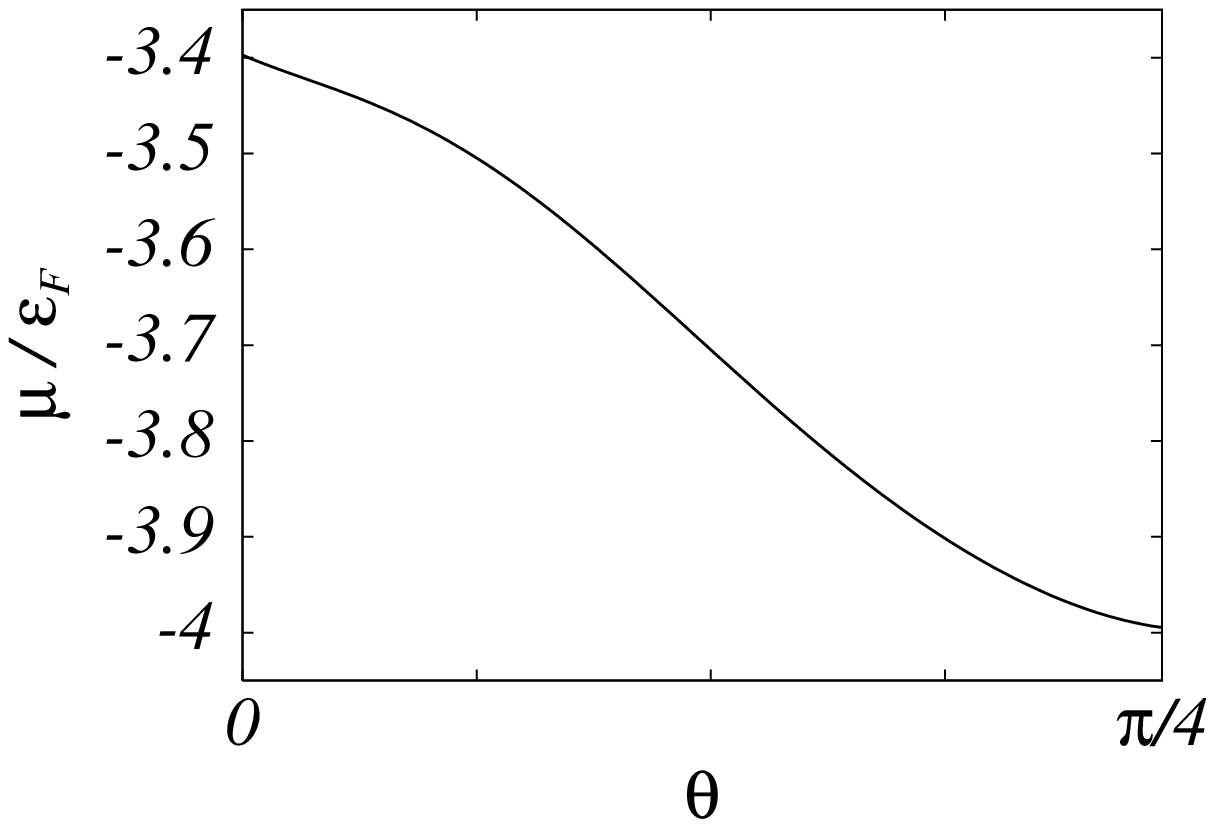}
\includegraphics[width=4.cm,height=4.cm]{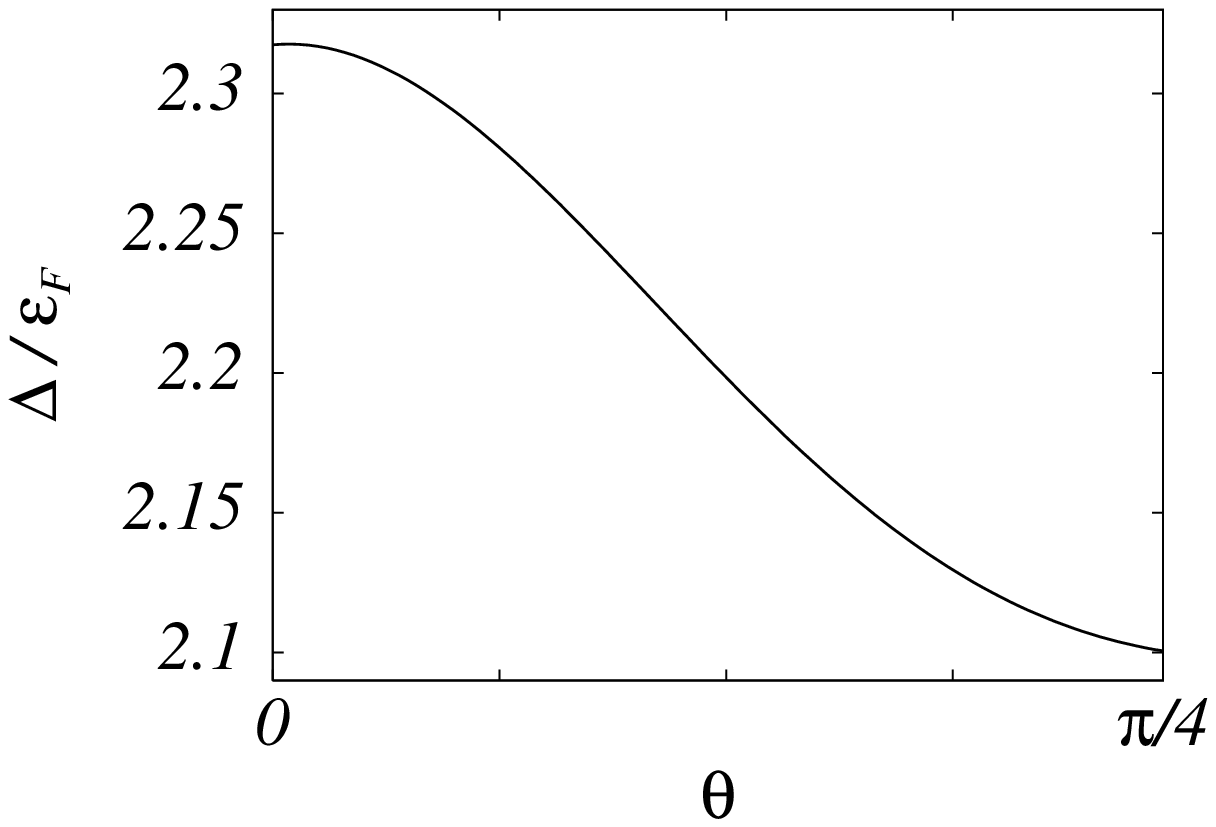}
\caption{(Color online) 2D Fermi superfluid.
Top panel: Singlet (red solid line, left y-axis) and triplet (green dashed
line, right y-axis) contributions to the condensate fraction,
for $v^2=2v_F^2$ and $\epsilon_B=2 \epsilon_F$, as
functions of $\theta$.
Left bottom panel: chemical potential $\mu$ (in units of $\epsilon_F$)
for $v^2=2v_F^2$ and
$\epsilon_B=2\epsilon_F$, as a function of $\theta$.
Right bottom panel: gap function, $\Delta$ (in units of $\epsilon_F$),
for $v^2=2v_F^2$ and
$\epsilon_B=2\epsilon_F$, as a function of $\theta$.}
\label{fig.4}
\end{figure}

In Figs. \ref{fig.3} and \ref{fig.4} we plot these quantities by choosing
$v^2=2v_F^2$ in correspondence to two different values of the binding energy.
These two binding energies are those for which a softening and a hardening of
the triplet contribution ($2n_1/n$) against $\theta$ is expected; thus
Fig. \ref{fig.3} is obtained with $\epsilon_B=0.5 \epsilon_F$ and
Fig. \ref{fig.4} with $\epsilon_B=2\epsilon_F$ (see the above discussion about
Fig. \ref{fig.1}).
From the former figure, Fig. \ref{fig.3}, it can be pointed out
that, for small binding energies, $2n_1/n$ decreases by
increasing $\theta$ from $0$ to $\pi/4$,
and the same behavior, but rescaled, is observed for the singlet
contribution $2n_0/n$ (see the top panel). Both the chemical potential
and the gap function decrease going from the only-Rashba case
to a fully mixed one.
Fig. \ref{fig.4}, instead, shows that, for large binding energies, $2n_1/n$
increases when $\theta$ is changed from $\theta=0$ to $\theta=\pi/4$,
while $2n_0/n$ always decreases, as well as the chemical potential and
the gap function.
In both above discussed cases, the singlet contribution
to the condensate fraction is always greater than the triplet one, as one can 
easily see from Eqs. (\ref{n0}) and (\ref{n1}).

In order to qualitatively explain the different behaviors of the two 
contributions to the condensate at different energies and spin-orbital mixings,
 it is important to consider the dispersions $E_1(\vk)$ and $E_2(\vk)$. 
In particular, since $E_2\ge E_1$, the main contributions to the sums in 
Eqs. (\ref{n0}) and (\ref{n1}) are due to the momenta at which $E_1(\vk)$ is 
minimum. Let us consider for simplicity the 2D case and rescale, for 
convenience, all the parameters in terms of the corresponding Fermi values: 
$\tilde \vk=\vk/k_F$, $\tilde v=v/v_F$, $\tilde\mu=\mu/\epsilon_F$, 
$\tilde \Delta=\Delta/\epsilon_F$, and ($i=1,2$)
\[
\widetilde{E}_i(\tilde \vk)=\sqrt{\left(\tilde{\vk}^2-\tilde\mu\mp 2\tilde v \sqrt{\tilde{\vk}^2+(\tilde{k}_y^2-\tilde k_x^2)\sin 2\theta}\right)^2+\tilde\Delta^2}.
\]
Notice that both $\tilde\mu$ and $\tilde\Delta$ depend on $\theta$, as shown in
the lower panels of Fig. \ref{fig.3} and Fig. \ref{fig.4}. 
Let us now focus on the minima of $\widetilde E_1$. At $\theta=\pi/4$, for 
instance, if $2\tilde v^2> -\tilde \mu$, 
the points in momentum space which minimize $\widetilde E_1$ belong to the 
two circumferences centered at $(0,\pm \tilde v)$ with radius $\sqrt{2\tilde 
v+\tilde\mu}$ (where $\xi_\vk-|\gamma(\vk)|=0$), so that 
$\widetilde E_{1 min}=\tilde \Delta$. This seems to be the case for low 
scattering strengths, at which those momenta mainly contribute to $n_0$ and 
$n_1$ as shown in the right top panels of Fig. \ref{fig.5} for the singlet 
and Fig. \ref{fig.6} for the triplet densities, 
at $\epsilon_B=0.5 \epsilon_F$ and $\tilde v=\sqrt{2}$, where the two rings 
are clearly highlighted. However, as shown in those figures, the most 
important contributions are due to 
$\pm \left(0, \sqrt{2}\,\tilde v+\sqrt{\tilde \mu+2\tilde v^2}\right)$ 
for the singlet and $\left(0, \sqrt{2}\,\tilde v
-\sqrt{\tilde \mu+2\tilde v^2}\right)$ for the triplet, because of the 
relative sign appearing in Eqs. (\ref{n0}) and (\ref{n1}). 
Quite in general, because of the monotonicity of $\widetilde E_2(\tilde\vk)$ in momentum space, 
the main contributions to the two condensates for finite $\theta$ are due to: $i$) momenta close to
\bean
&&\hspace{-0.3cm}
\tilde \vk_1=\pm\left(0, \tilde v \sqrt{1+\sin 2\theta}-\sqrt{\tilde \mu+
\tilde v^2(1+\sin 2\theta)}\right),\;\; \textrm{for}\; n_0,\\
&&\hspace{-0.3cm}
\tilde \vk_1=\pm\left(0, \tilde v \sqrt{1+\sin 2\theta}+\sqrt{\tilde \mu+
\tilde v^2(1+\sin 2\theta)}\right),\;\; \textrm{for}\; n_1,
\eean
if $\tilde v^2(1+\sin 2\theta)> -\tilde \mu$, and $ii$) momenta close to 
\[\tilde \vk_2=\pm\left(0, \tilde v \sqrt{1+\sin 2\theta}\right),\] 
for $\tilde v^2(1+\sin 2\theta)\le -\tilde \mu$. For $\theta=0$ the condition 
$\tilde v^2\le -\tilde \mu$ occurs almost always, 
except for small $v$ and low binding energies, as one can check 
by looking at Fig. 6 of Ref. \cite{dms}, therefore, for only Rashba (or only 
Dresselhaus) spin orbit coupling, since rotational symmetry is recovered, 
the relevant momenta is distributed almost always around a single ring 
ceneterd at $\vk=0$ and radius $\tilde v$, widened inwards for $n_0$ and 
outwards for $n_1$ due to the monotonicity of $E_2(\vk)$ (see first top and 
bottom panes in Figs. \ref{fig.5}, \ref{fig.6}). Only for sufficiently 
small $v$ and 
$\epsilon_B$, the relevant momenta are close to two concentric circles with 
radii $\tilde v\pm \sqrt{\tilde v^2+\tilde\mu}$. 
By mixing Rashba and Dresselhaus spin orbital couplings, therefore, we can 
filter particle pairs with relative wavevectors maily 
along a definite direction, 
excluding all the rest from partecipating to the condensate. 

This is the main reason of the suppression of $n_0$ and $n_1$, for low 
scattering strengths, see top panel of Fig. \ref{fig.3}. 
Moreover, for low binding energy, the low lying energy level is 
\[\tilde E_{1 min}=\tilde E_1(\tilde\vk_1)=\tilde \Delta,\]  
for $\theta> \theta^*$, 
where $\theta^*$ is such that $\tilde v^2(1+\sin 2\theta^*)+\tilde \mu=0$ (for $\tilde v=\sqrt{2}$ and $\tilde\epsilon_B=0.5$, as in top panels of Figs. \ref{fig.5}, \ref{fig.6}, $\theta^*\simeq 0.1$). Since in Eqs. 
(\ref{n0}) and (\ref{n1}) the main quantity is ${\Delta}/{E_1}$, 
for $\theta >\theta^*$ the maximum values of the singlet and triplet densities 
weakly depend on $\theta$, as shown by the top rows of plots in Figs. \ref{fig.5}, \ref{fig.6}. Therefore, even if the top signal in both $|\langle\psi_\uparrow(\vk) \psi_\downarrow(-\vk)\rangle|$ and $|\langle\psi_\uparrow(\vk) \psi_\uparrow(-\vk)\rangle|$ remains basically the same for different $\theta$'s, the condensate is suppressed by downsizing the wavevector domain (from a broad 
large ring for $\theta=0$, to only two spots for $\theta=\pi/4$), or in other 
words, by reducing the degrees of freedom of the particle pairs. 

On the contrary, at large scattering parameters, the competition of several 
effects play a role. 
Also for large scattering, the mixing of the two spin orbital coupling 
reduces the domain of relevant momenta from a broad large ring to two spots, 
but, at the same time, the effective gap in the spectrum is reduced. 
Within the set of parameters used, the energy gap is always greater than the 
pairing function $\Delta$, and is 
\[\tilde E_{1 min}= \tilde E_1(\tilde\vk_2)=
\sqrt{\tilde\Delta^2+\left(\tilde v^2(1+\sin 2\theta)+\tilde \mu\right)^2},\]
which, moreover, decreases faster than $\tilde\Delta$, upon 
increasing $\theta$. 
Notice, by the way, that at $\theta =\pi/4$, because of Eqs. (\ref{erd_mu}), 
(\ref{erd_del}),  
one obtain the same gap as without spin-orbit interaction, 
$\sqrt{\tilde\Delta^2+\tilde \mu^2}\,\big{|}_{v=0}$. 
The increase of the intensity of the condensate densities, due to the increase 
of ${\Delta}/{E_{1min}}$ with $\theta$, competes with the wavevector domain 
reduction. It is crucial, at this point, to study the behavior of the second 
branch of the spectrum, i.e. $\tilde E_2(\tilde\vk)$, which, at the points 
where $\tilde E_1$ is minimum, is given by
\[\tilde E_2(\tilde\vk_2)=
\sqrt{\tilde\Delta^2+\left(\tilde 3v^2(1+\sin 2\theta)-\tilde \mu\right)^2}.\] 
The second branch, contrary to $\tilde E_{1 min}$, is an increasing function 
of $\theta$, therefore 
it tries to suppress the singlet condensate while promoting the triplet one. 

\begin{widetext}
\phantom{.}
\begin{figure}[ht]
\begin{tabular}{cccc}
$\theta=0$ & $\theta=\pi/16$ & $\theta=\pi/8$ & $\theta=\pi/4$\\
\includegraphics[width=4.cm]{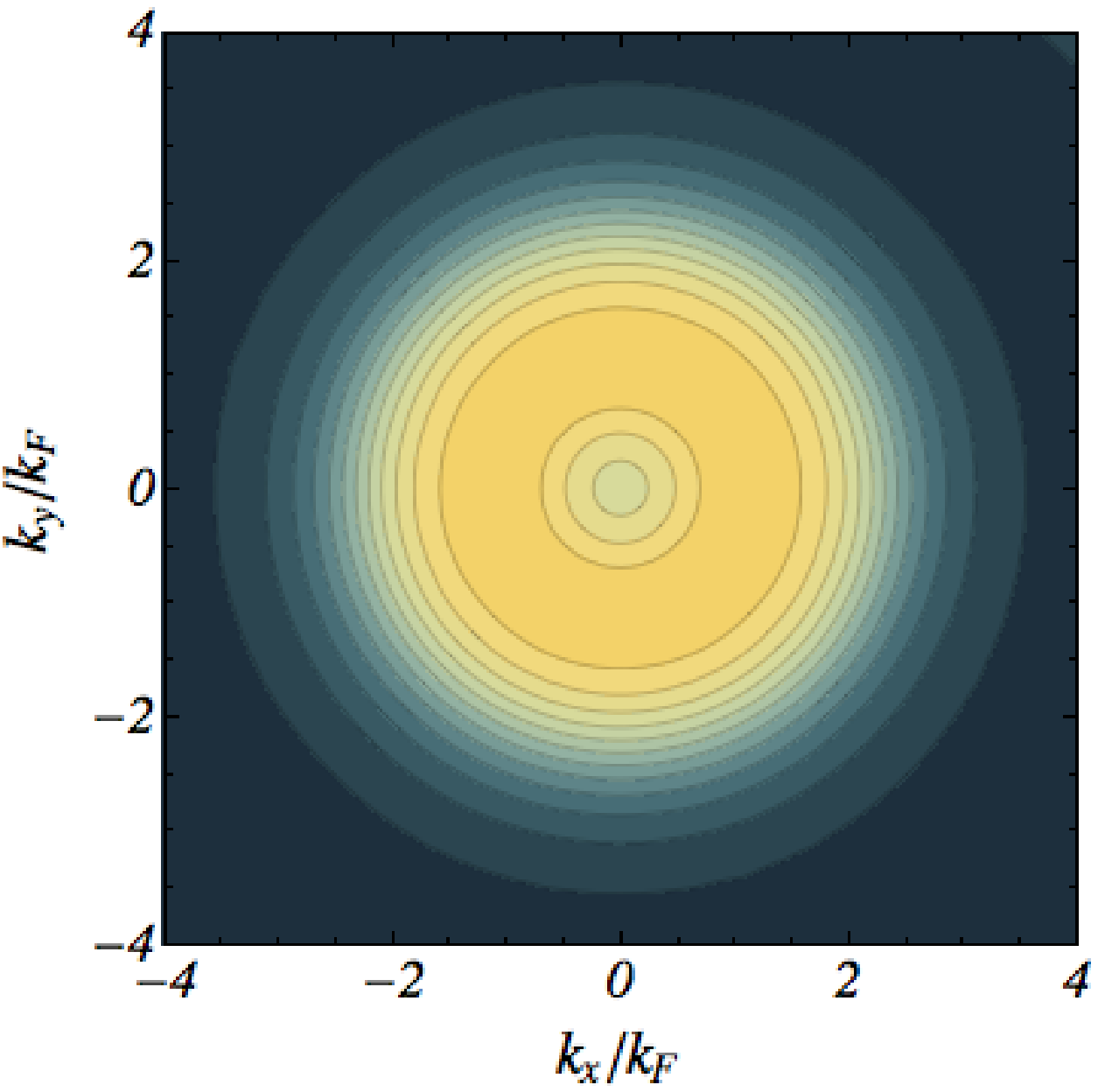}&
\includegraphics[width=4.cm]{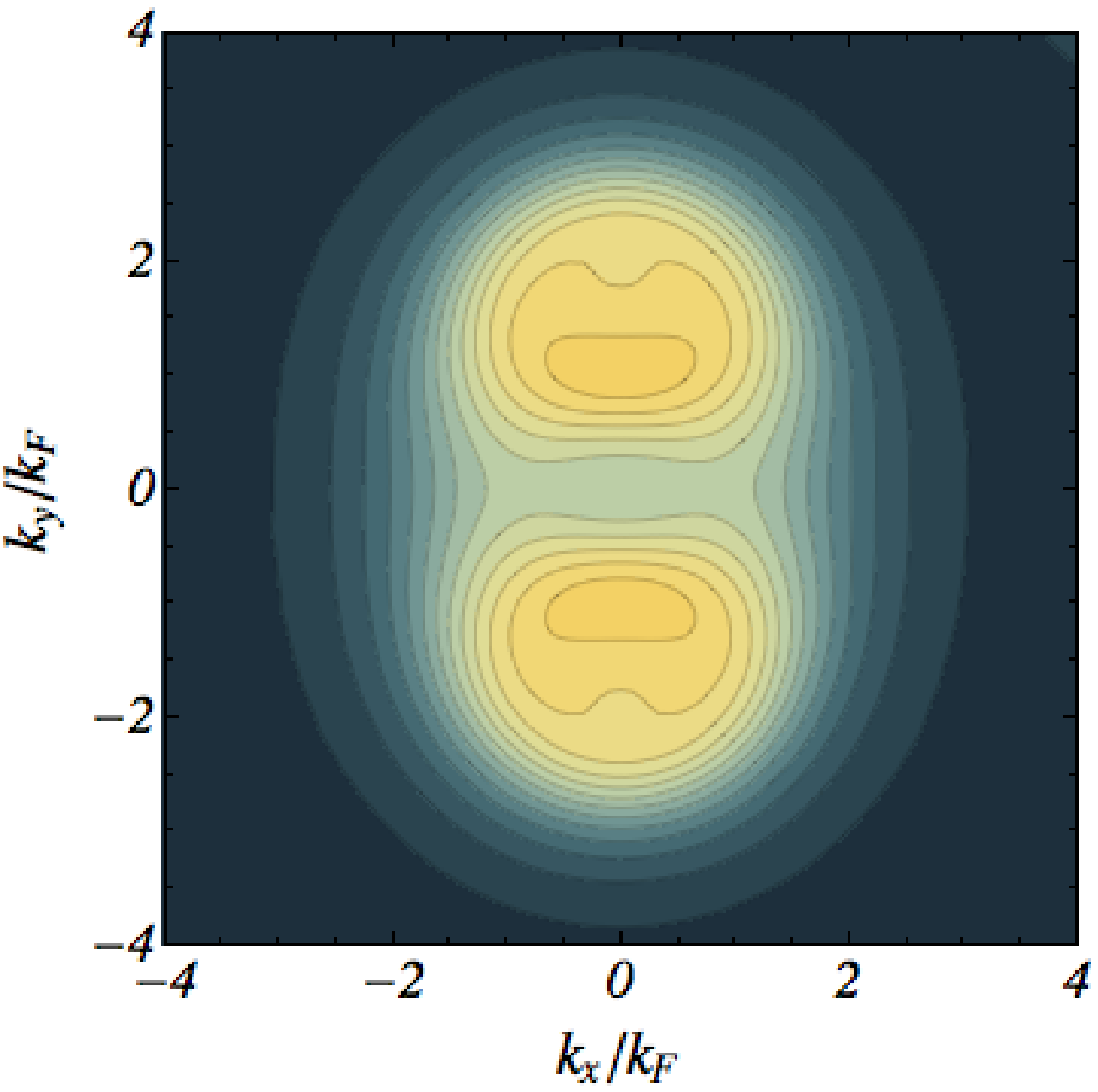}&
\includegraphics[width=4.cm]{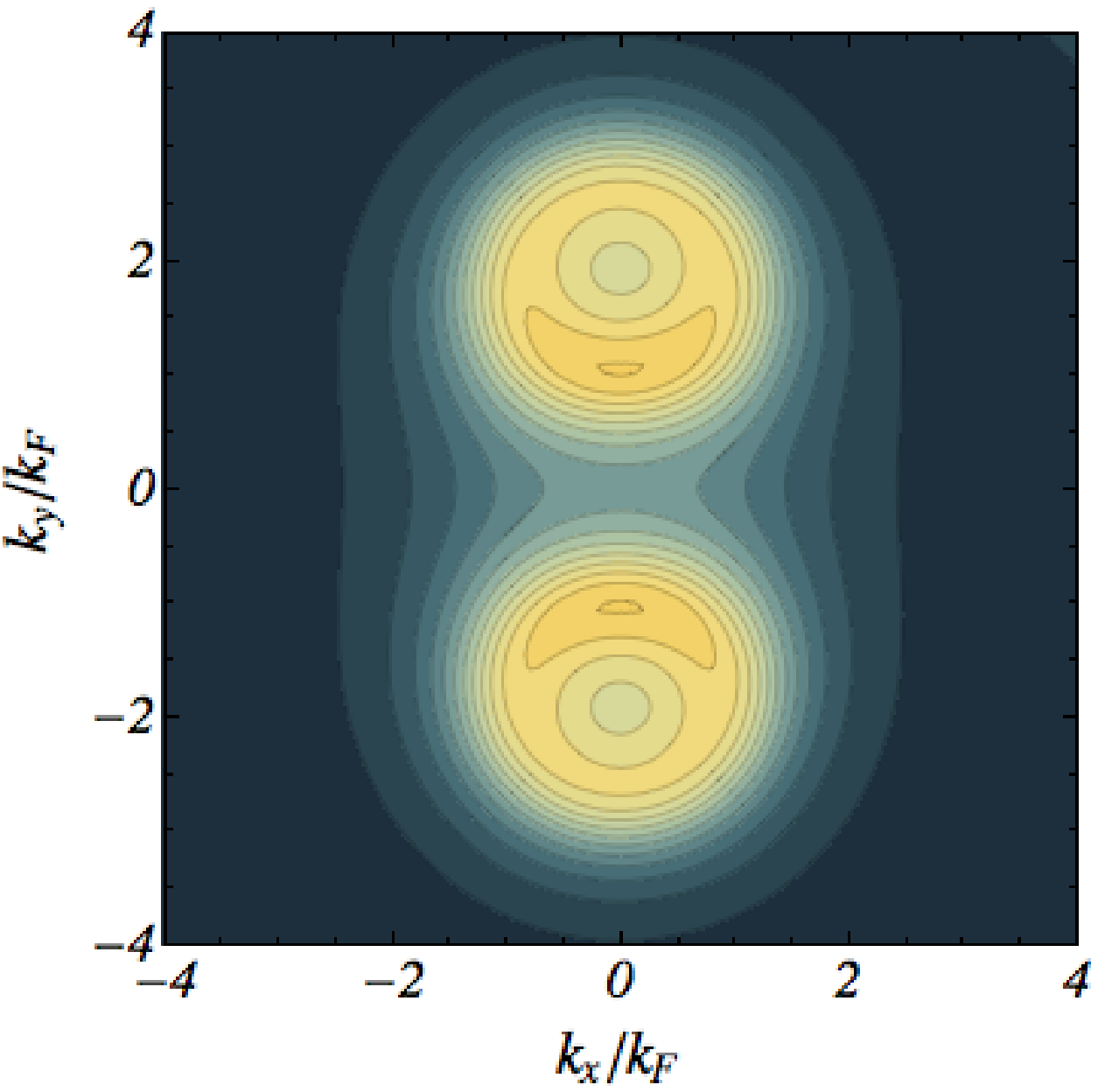}&
\includegraphics[width=4.cm]{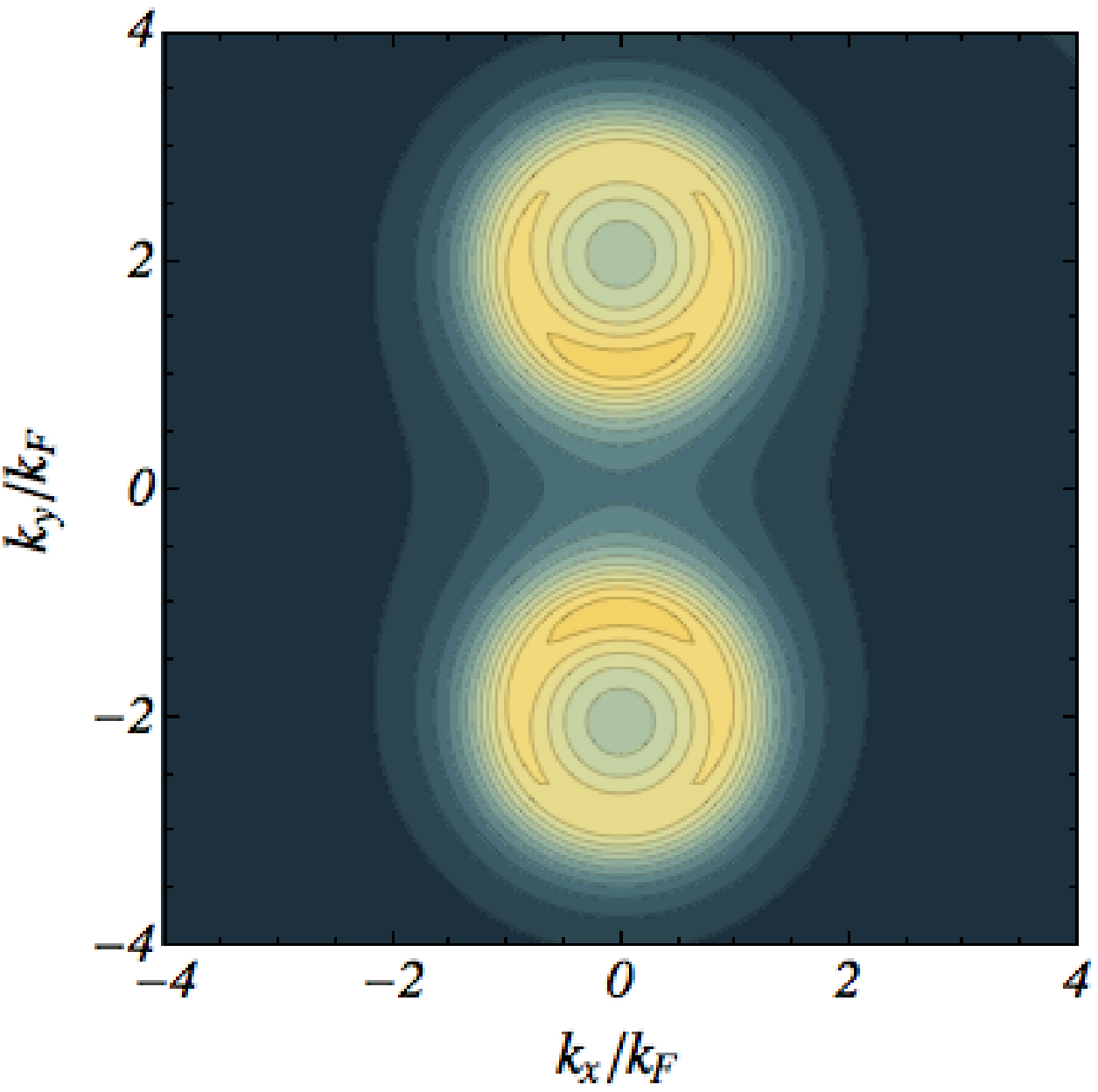}\\
\includegraphics[width=4.cm]{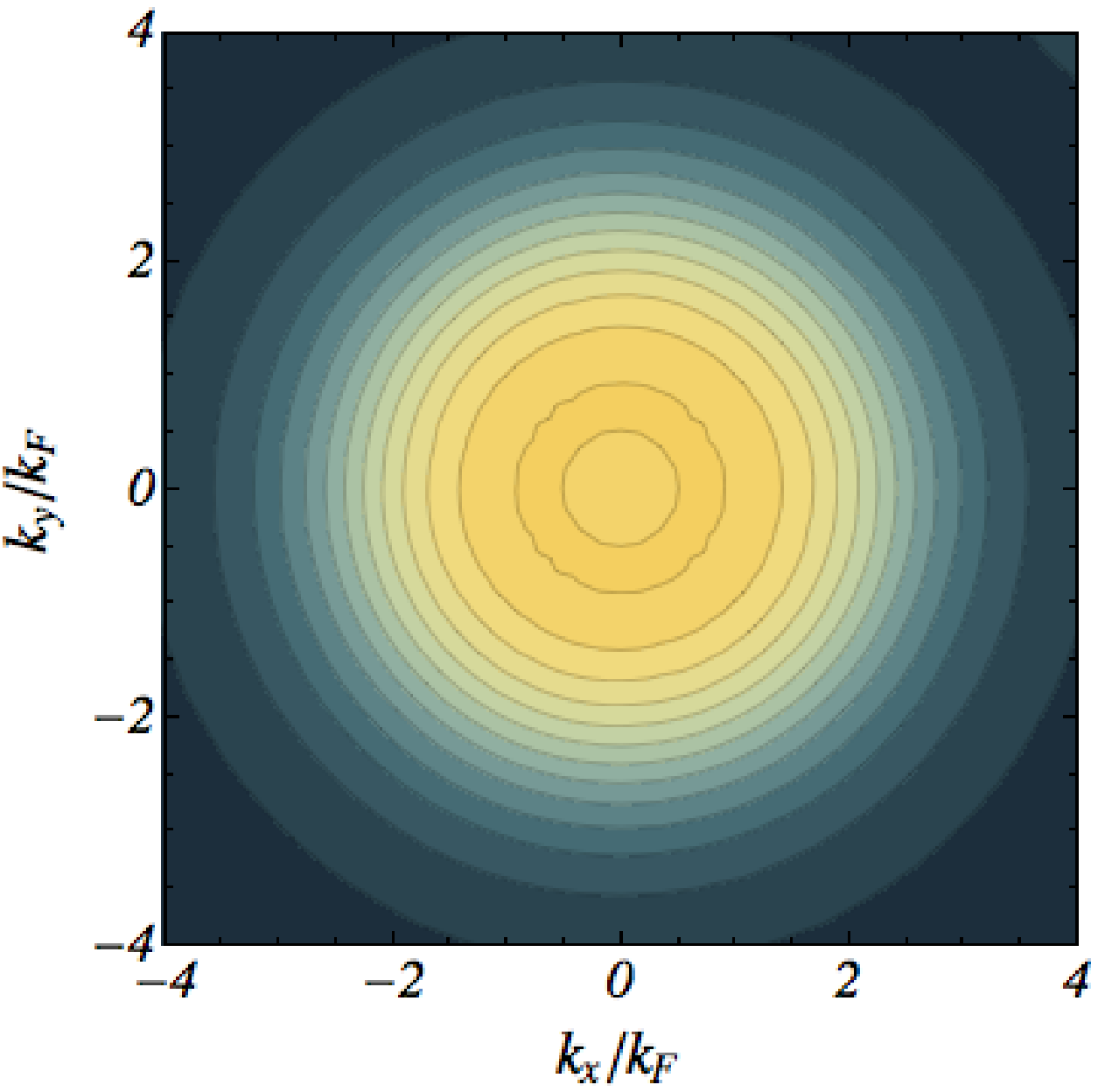}&
\includegraphics[width=4.cm]{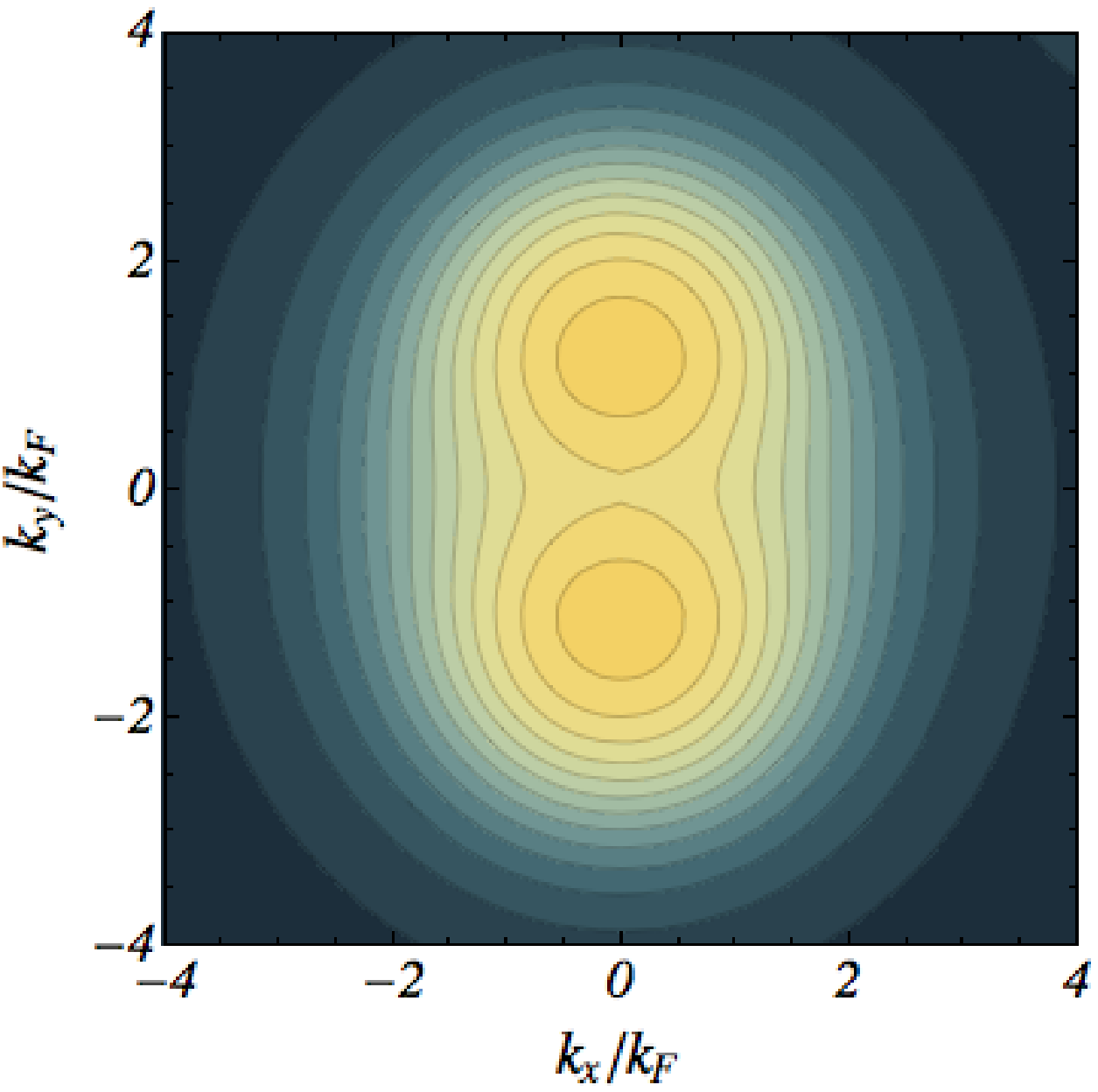}&
\includegraphics[width=4.cm]{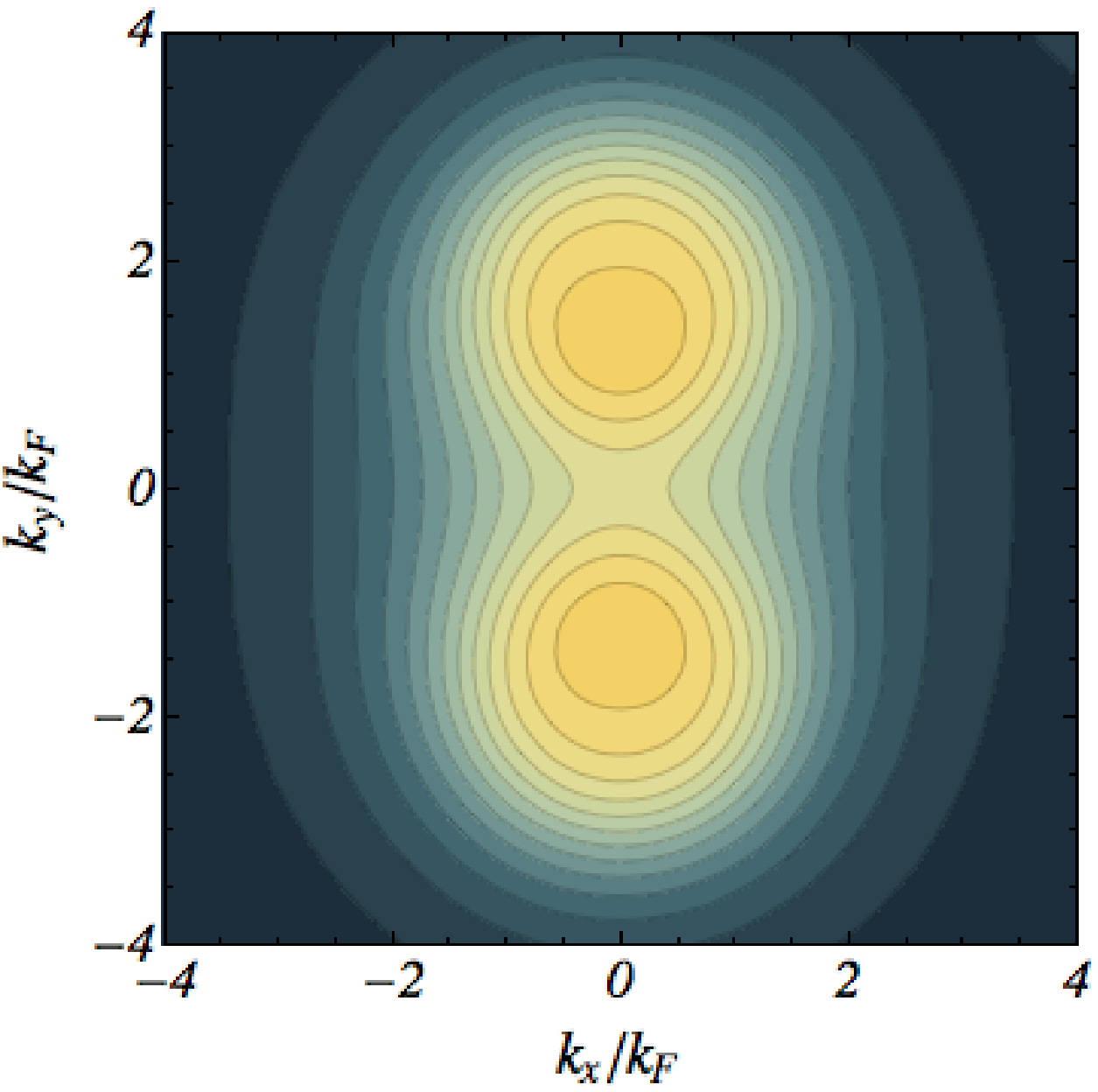}&
\includegraphics[width=4.cm]{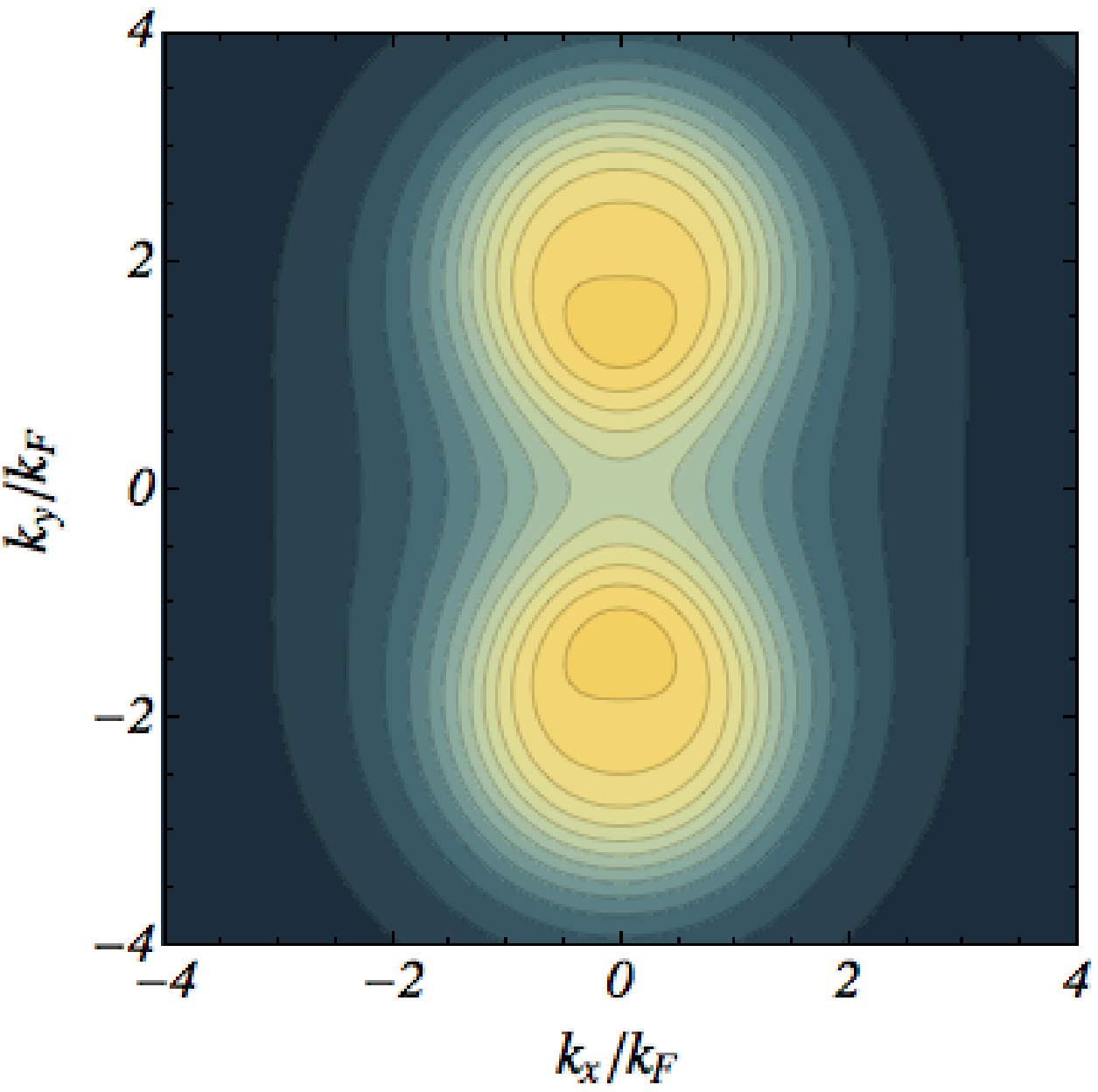}
\end{tabular}
\caption{(Color online) 2D Fermi superfluid. Contourplot of the singlet pairing
$\frac{1}{(2\pi)^2 n}|\langle \psi_\uparrow(\vk) \psi_\downarrow(-\vk)\rangle|^2$ in momentum space (as a function of the rescaled dimensionless momenta $\vk/k_F$), for $v=\sqrt{2}v_F$ and different values of $\theta$, ($\theta=0, \pi/16, \pi/8, \pi/4$ from left to right) and for $\epsilon_B=0.5 \epsilon_F$ (upper plots), $\epsilon_B=2 
\epsilon_F$ (lower plots). The brighter the higher is the value of the singlet 
density, from $0$ (deep blue) to $0.003$ (intense yellow). The integral over 
the dimensionless momenta $\vk/k_F$, gives $2n_0/n$ as in Fig. \ref{fig.3} 
(for $\epsilon_B=0.5 \epsilon_F$) and Fig. \ref{fig.4} (for $\epsilon_B=2 \epsilon_F$).}
\label{fig.5} 
\end{figure}
\begin{figure}[ht]
\begin{tabular}{cccc}
$\theta=0$ & $\theta=\pi/16$ & $\theta=\pi/8$ & $\theta=\pi/4$\\
\includegraphics[width=4.cm]{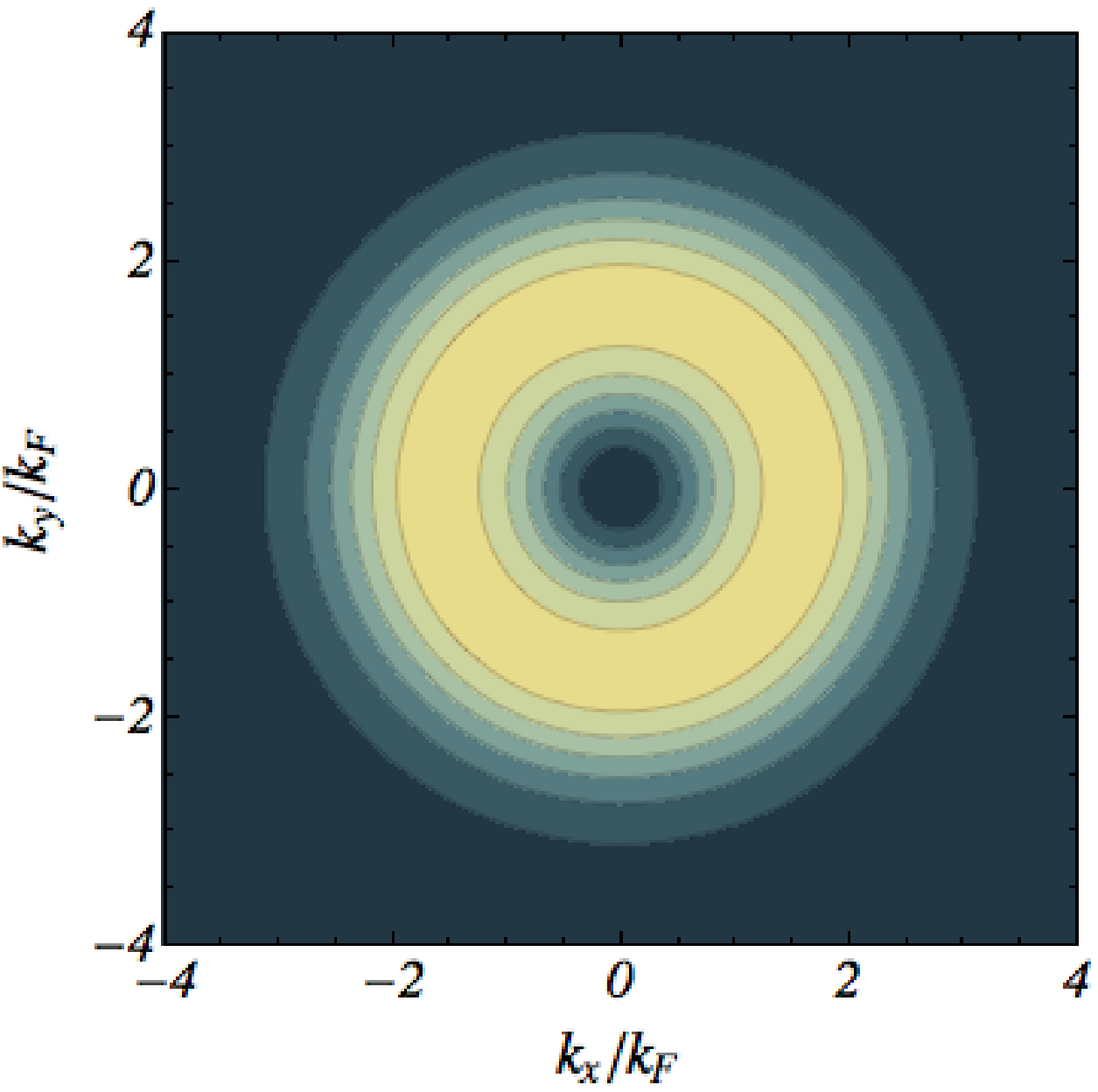}&
\includegraphics[width=4.cm]{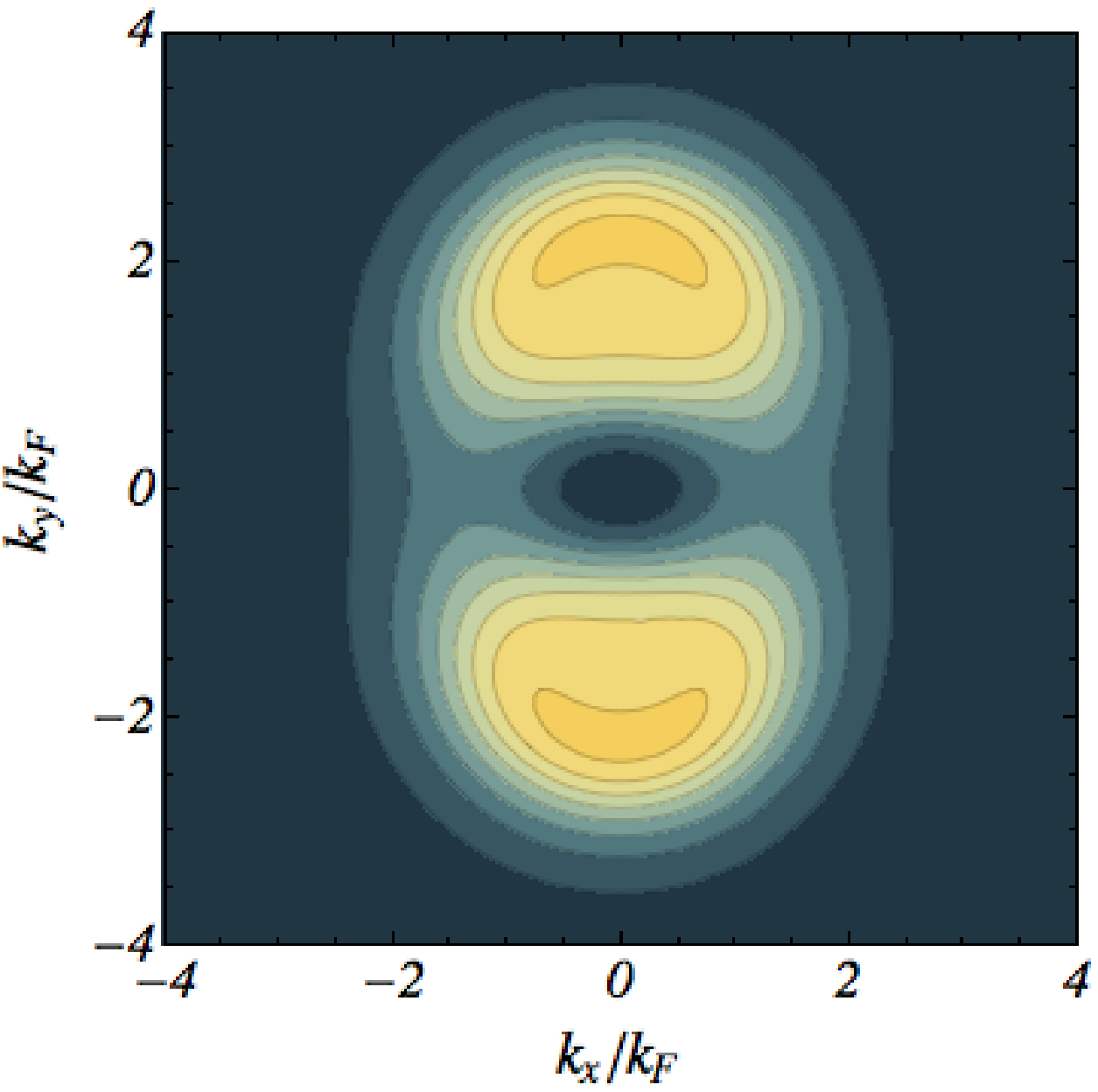}&
\includegraphics[width=4.cm]{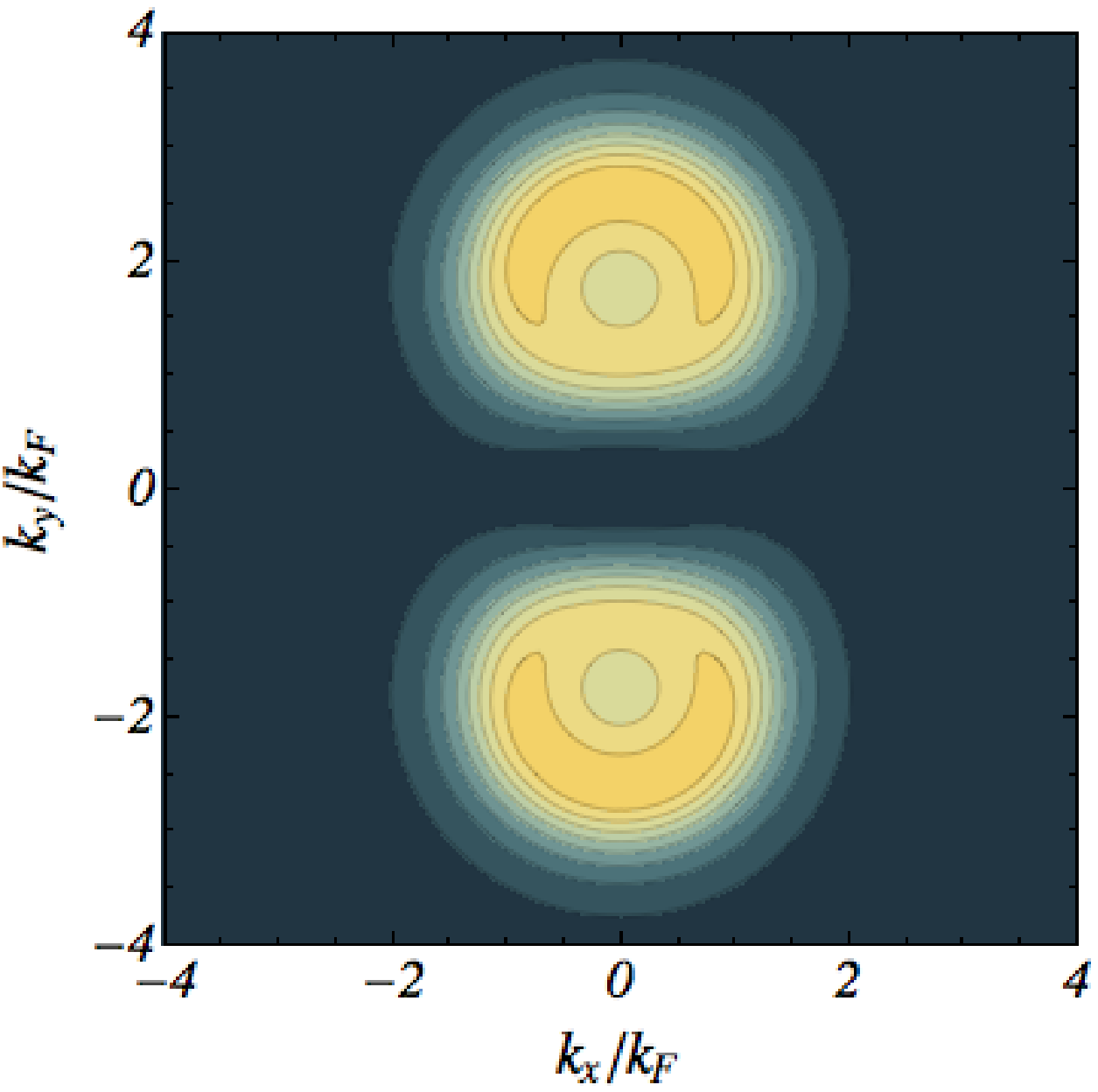}&
\includegraphics[width=4.cm]{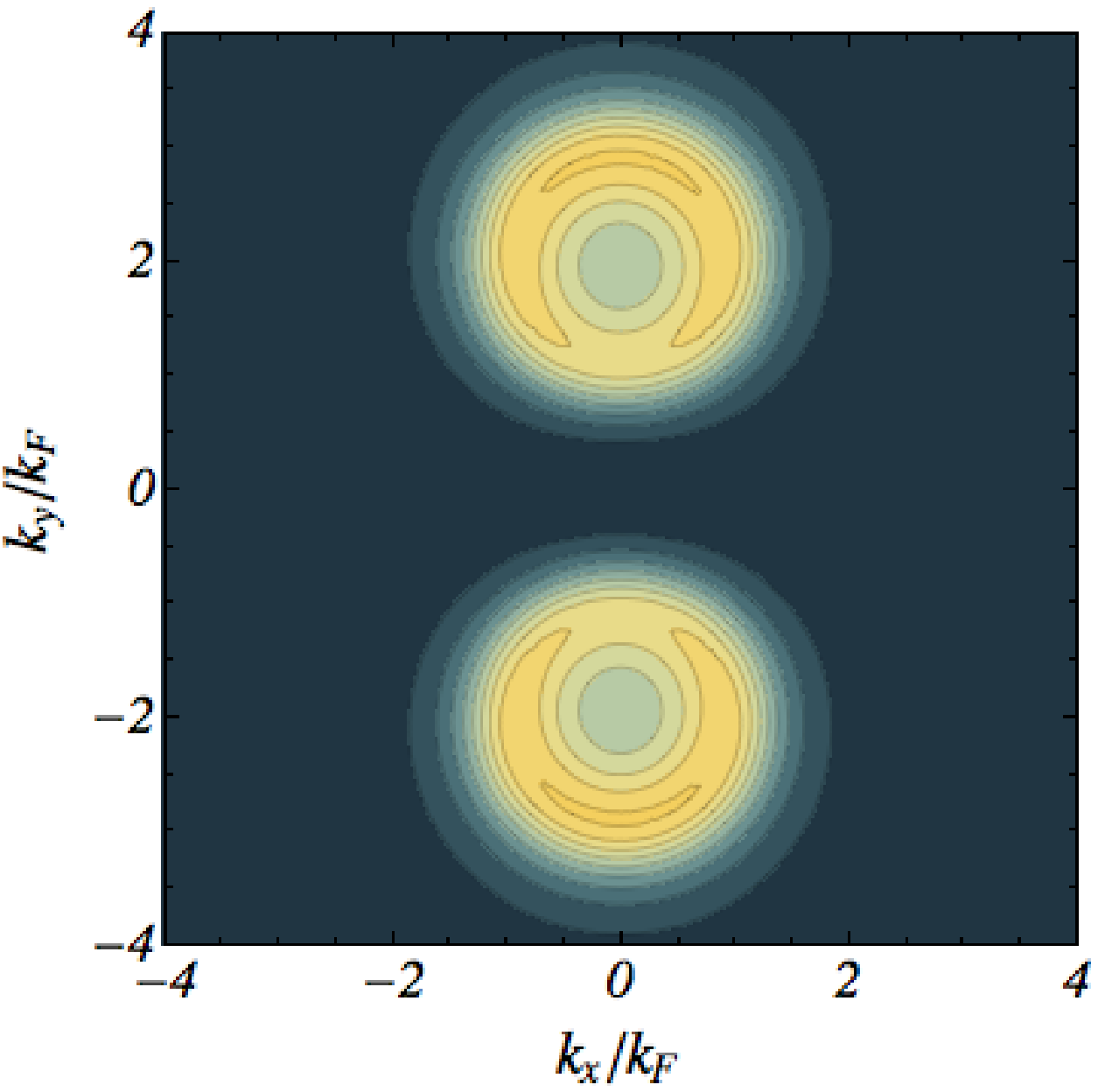}\\
\includegraphics[width=4.cm]{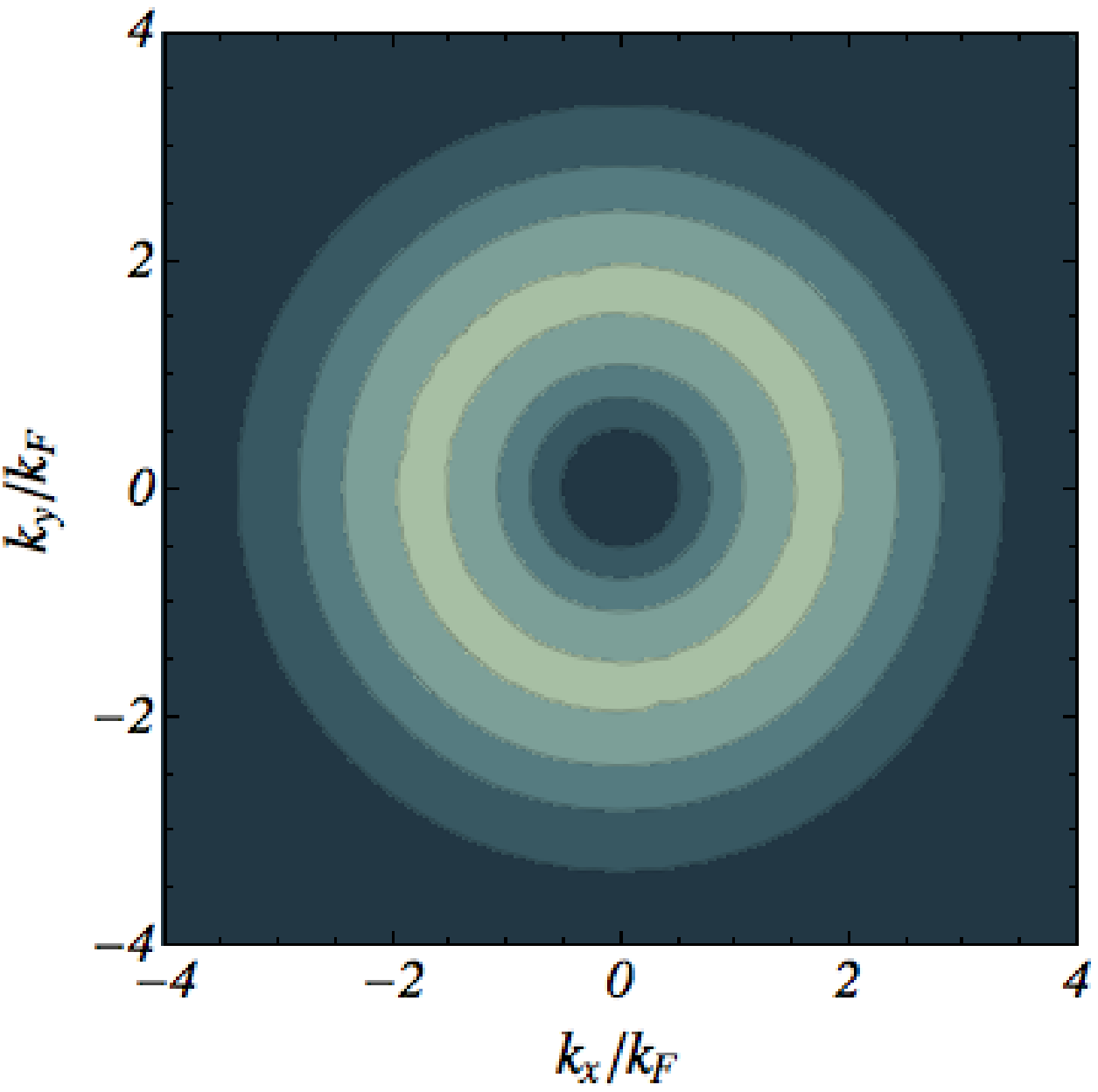}&
\includegraphics[width=4.cm]{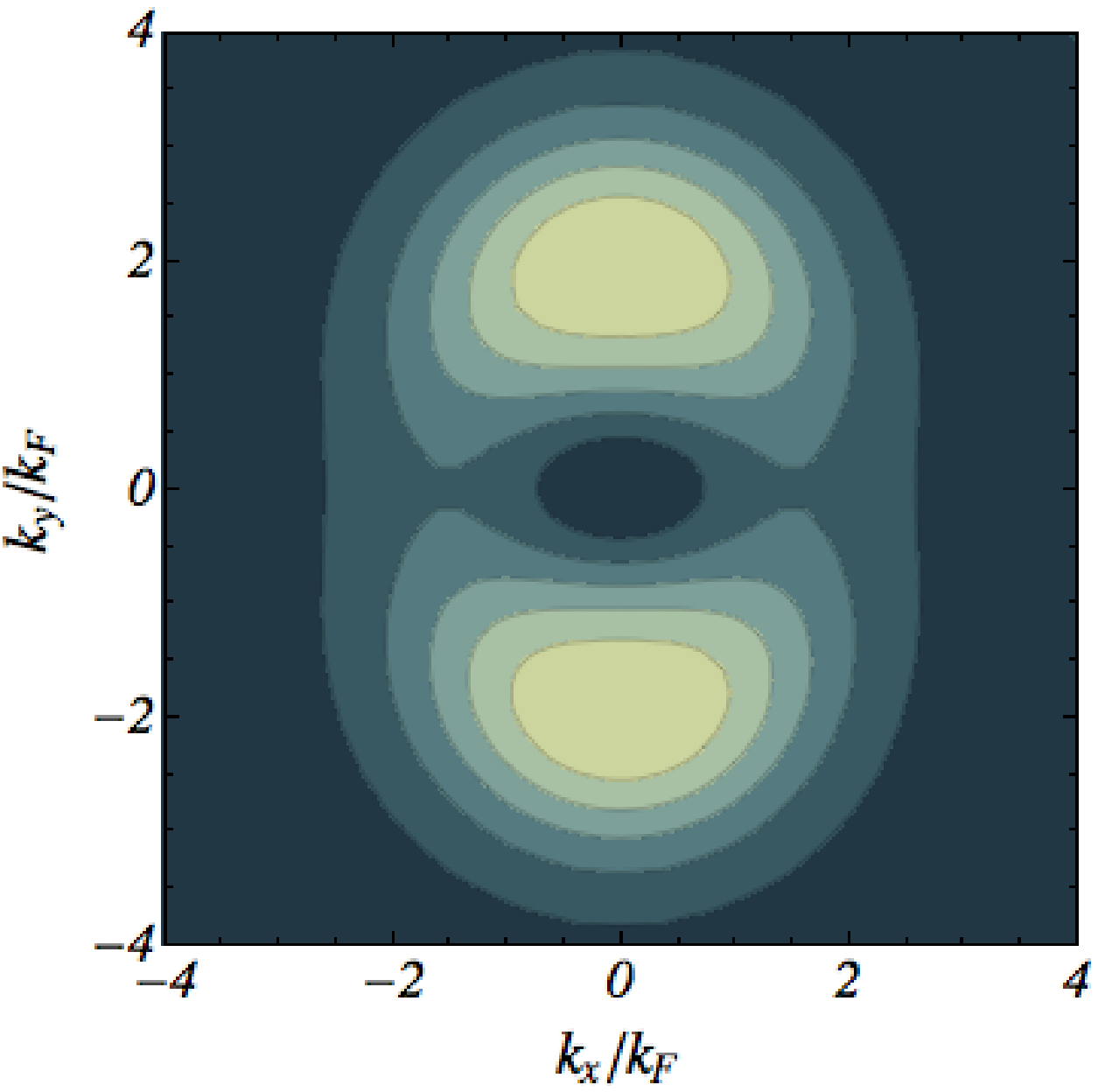}&
\includegraphics[width=4.cm]{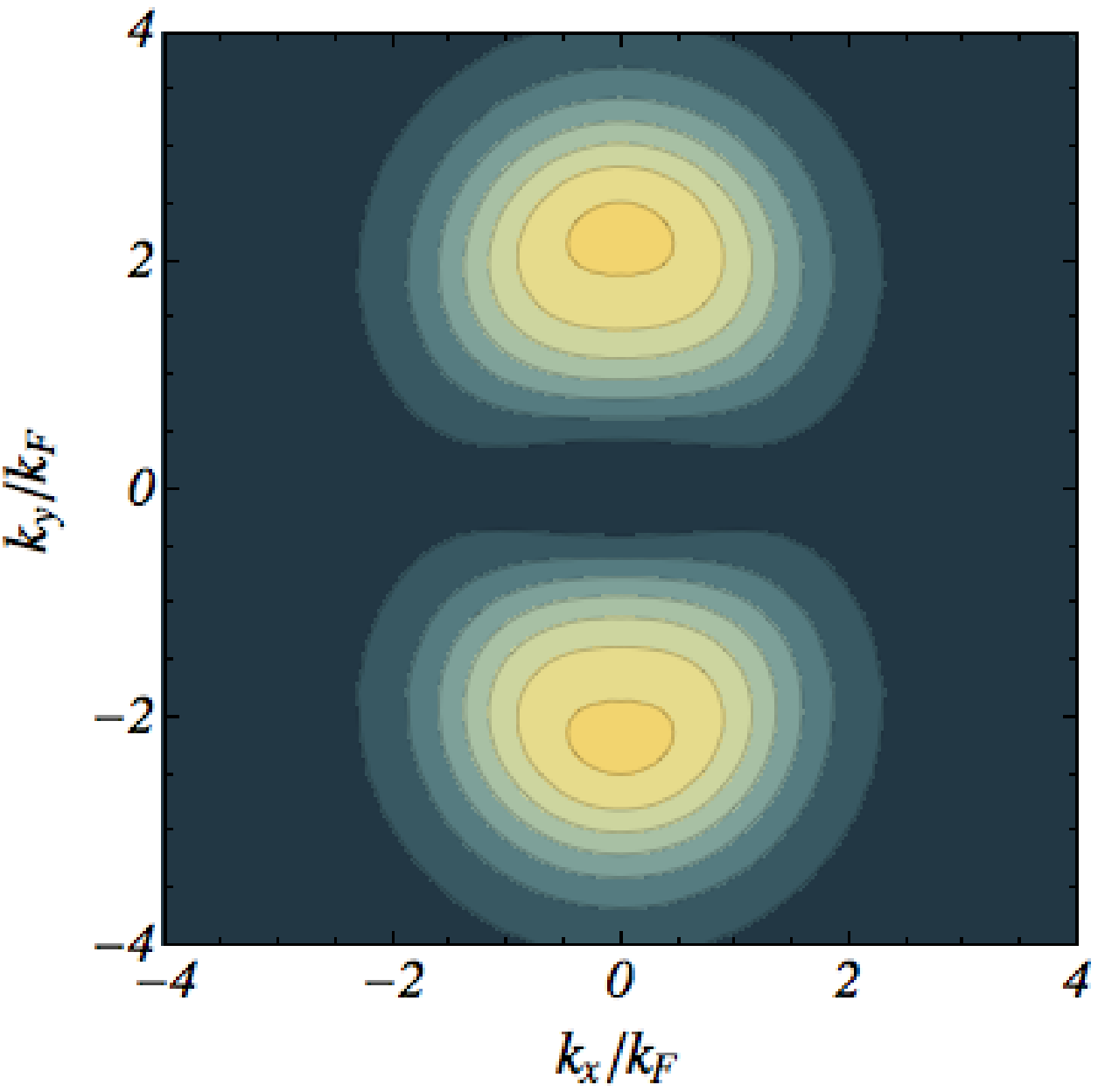}&
\includegraphics[width=4.cm]{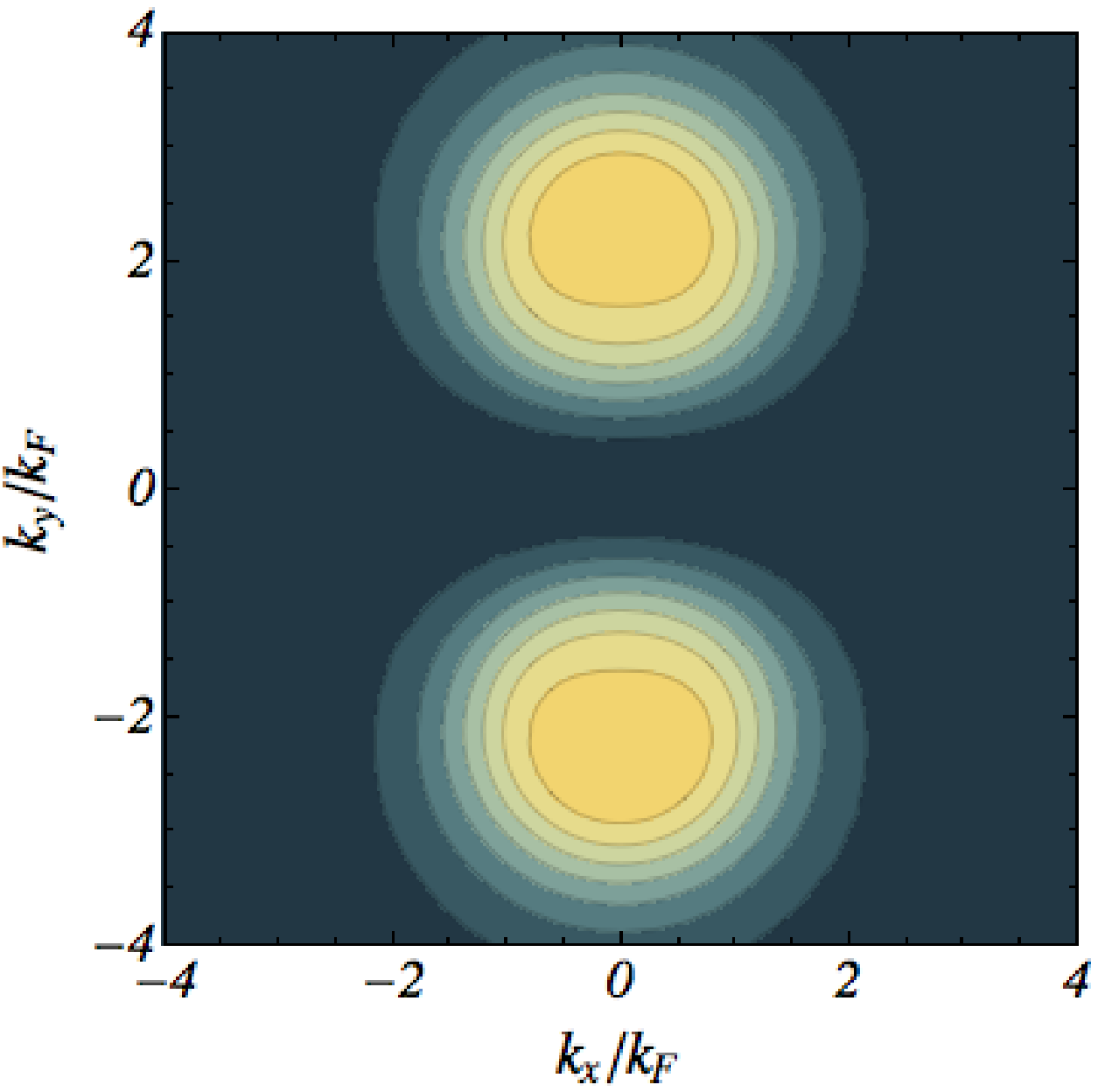}
\end{tabular}
\caption{(Color online) 2D Fermi superfluid. Contourplot of the triplet pairing
 $\frac{1}{(2\pi)^2 n}|\langle \psi_\uparrow(\vk) \psi_\uparrow(-\vk)\rangle|^2$ in momentum 
space, for $v=\sqrt{2} v_F$ and different values of $\theta$, ($\theta=0, \pi/16, \pi/8, \pi/4$ 
from left to right) and for $\epsilon_B=0.5 \epsilon_F$ (upper plots), $\epsilon_B=2\epsilon_F$ (lower plots). 
The brighter the higher is the value of the triplet density, from $0$ (deep blue) to $0.002$ (intense yellow). The integral over the dimensionless momenta $\vk/k_F$ gives $2n_1/n$ as in Fig. \ref{fig.3} (for $\epsilon_B=0.5 \epsilon_F$) and Fig. \ref{fig.4} (for $\epsilon_B=2 \epsilon_F$).}
\label{fig.6}
\end{figure}

\end{widetext}

In the Introduction we have claimed that by extending the procedure used
in previous experiments \cite{zwierlein,ueda} one can measure the condensate fraction of singlet and triplet pairs. The procedure that we suggest is as follows. In the BEC side of the 3D crossover
(or in the full 2D crossover) one first applies a
Stern-Gerlach field gradient \cite{zwierlein} on the cloud
to spatially separate the
molecules in the $\pm 1$-triplet state from the rest. Supposing 
equipartition of the three triplet components one can count the total number 
of molecules in the triplet state and in the singlet one. 
One measures, indeed, the momentum distribution
of each cloud from which the fraction of molecules in the
zero-momentum state is extracted. In this way one gets the condensate
fraction of molecules in singlet and triplet states \cite{zwierlein,ueda}.
Clearly, in the presence of a space-dependent trapping potential the condensed molecules are not in a zero-momentum state but in a state with a finite width (in the momentum space) \cite{zwierlein,ueda} which depends on the choice of the confining potential.
In the BCS side of the 3D crossover the procedure is slightly different. In this case one wants to measure the condensate fraction of Cooper pairs which are not in a true bound state. {The key point is to apply a magnetic field ramp adiabatic with respect to two-body physics but fast with respect to many body physics in such a way to transfer} Cooper pairs of atoms into bound molecules \cite{zwierlein}. After that one uses
a Stern-Gerlach field gradient \cite{zwierlein} to spatially separate the molecules (with spin zero and one) and atoms (with spin one half). Finally,
from the momentum distribution of molecules one deduces
the condensate fractions (singlet and triplet) of the initial BCS state.

\section{Conclusions}

We have analyzed the condensation of fermionic atoms
along the BCS-BEC crossover in the presence of Rashba and
Dresselhaus spin-orbit couplings. The condensation has been characterized
by calculating the singlet and the triplet contributions
to the pairing and therefore the
full condensate fraction. We have studied these quantities by varying
the spin-orbit from the situation in which the only Rashba coupling is
present to that in which the Rashba and Dresselhaus velocities
are equal. We have found that moving along this path, the singlet
contribution to the condensate fraction decreases, while the triplet
one behaves differently in the two regimes (BCS and BEC).
In the BCS regime, the triplet pairing is suppressed upon mixing the two
spin-orbit couplings while in the BEC regime it experiences an enhancement
over the only-Rashba case.
In other words, the triplet pairing is maximized in the BCS regime if
only Rashba (or only Dresselhaus) term is active, while it is strengthened
in the BEC regime by mixing the two spin-orbital couplings.
This behavior takes place both in two and three dimensions and can be 
explained by studing the properties of the spectrum. In the BCS regime the 
dominant effect of the spin-orbital mixing is the selection of particle pairs 
by a wavevector filtering, reducing the number of those which participate to 
the condensate. On the BEC regime, instead, several effects can compete or 
cooperate upon increasing the Rashba-Dresselhaus mixture: 
the momentum domain reduction, the decrease of the energy gap and the 
increase of the steepness of the second branch of the spectrum, which finally 
can suppress the singlet condensate promoting the triplet one. 
We have shown also that the total condensate fraction 
is greater when only one coupling (only Rashba or only Dresselhaus)
is present, while in the equal-Rashba-Dresselhaus case,
is the same as that obtained without spin-orbit. 
{Finally, we have suggested that
the condensate fraction of singlet and triplet pairs
may be detected by suitably extending the experimental procedures employed
in Refs. \cite{zwierlein,ueda}}.

\end{document}